\documentclass[12pt,a4paper]{iopart}

\usepackage{iopams}
\usepackage{setstack}
\usepackage{graphicx}
\usepackage{pict2e}
\usepackage{color}
\usepackage{hyperref}

\newcommand{\C}[2]{{#1 \choose #2}}
\newcommand{\Li}{\mathrm{Li}}
\newcommand{\Ai}{\mathrm{Ai}}

\newcommand{\Res}{\mathrm{Res}}
\renewcommand{\Re}{\mathrm{Re}}
\renewcommand{\Im}{\mathrm{Im}}
\newcommand{\openone}{{\bf1}}
\newcommand{\half}{\frac{1}{2}}
\newcommand{\thalf}{\tfrac{1}{2}}

\newcommand{\affiliation}[1]{\address{#1}}
\renewcommand{\pacs}[1]{\noindent\textbf{PACS numbers:} #1}
\newcommand{\keywords}[1]{\noindent\textbf{Keywords:} #1}
\newcommand{\tfrac}[2]{\mbox{\small$\frac{#1}{#2}$}}
\renewcommand{\text}[1]{\mathrm{#1}}
\renewcommand{\vec}[1]{{\bf #1}}

\begin{document}

\title[Current fluctuations and large deviations for TASEP on the relaxation scale]{Current fluctuations and large deviations for periodic TASEP on the relaxation scale}
\author{Sylvain Prolhac}
\affiliation{Laboratoire de Physique Th\'eorique; IRSAMC; UPS; Universit\'e de Toulouse; France\\Laboratoire de Physique Th\'eorique; UMR 5152; Toulouse; CNRS; France}

\begin{abstract} The one-dimensional totally asymmetric simple exclusion process (TASEP) with $N$ particles on a periodic lattice of $L$ sites is an interacting particle system with hopping rates breaking detailed balance. The total time-integrated current of particles $Q$ between time $0$ and time $T$ is studied for this model in the thermodynamic limit $L,N\to\infty$ with finite density of particles $\overline{\rho}=N/L$. The current $Q$ takes at leading order a deterministic value which follows from the hydrodynamic evolution of the macroscopic density profile by the inviscid Burgers' equation. Using asymptotics of Bethe ansatz formulas for eigenvalues and eigenvectors, an exact expression for the probability distribution of the fluctuations of $Q$ is derived on the relaxation time scale $T\sim L^{3/2}$ for an evolution conditioned on simple initial and final states. For flat initial and final states, a large deviation function expressed simply in terms of the Airy function is obtained at small rescaled time $T/L^{3/2}$.\\\\
\\\\
\end{abstract}


\maketitle

\begin{section}{Introduction}
\label{section introduction}
Lattice gases are interacting particle systems encountered in both equilibrium and non-equilibrium statistical mechanics. They are used as microscopic models for various physical and biological phenomena \cite{CMZ2011.1}. At large scales, considering macroscopic observables instead of the individual particles, these systems often evolve in time by deterministic hydrodynamic conservation laws. Understanding better fluctuations beyond the hydrodynamic behaviour is recognized as crucial in order to build a general theory for non-equilibrium phenomena \cite{S1991.1,D2007.1,BDSGJLL2015.1}. In many cases, the stochastic processes describing these fluctuations at large scale are independent of the details of the microscopic dynamics. This universal character of the fluctuations makes it very desirable to have exact expressions describing their statistics. This can be achieved by considering specific microscopic models simple enough so that they may be solved. This approach was successfully used in the past for equilibrium statistical mechanics, the Ising model being a notable example.

Another such model is the asymmetric simple exclusion process (ASEP) \cite{D1998.1,S2001.1,GM2006.1,D2007.1,S2007.1,M2011.1}, whose dynamics breaks detailed balance and has thus a true non-equilibrium steady state at stationarity. ASEP is known to be integrable in the sense of quantum integrability, also called stochastic integrability \cite{S2012.1} in the context of classical stochastic systems where convergence to a stationary state is ensured by the fact that the evolution operator is real valued, unlike in more traditional quantum integrable systems with unitary evolution where the issue of thermalization is still not completely settled.

It is usually possible to diagonalize exactly the evolution operators of integrable models for finite size systems using Bethe ansatz. For ASEP this leads, at least in principle, to exact expressions for the fluctuations. A technical problem is however to take the large scale limit of the finite size, finite time formulas, which is usually complicated as it involves delicate asymptotics of large determinants with entries written in terms of solutions of a large system of coupled polynomial equations of high degree. The situation simplifies enormously for the totally asymmetric simple exclusion process (TASEP), a special case of ASEP, for which some determinants can be computed explicitly, and the polynomial system of equations essentially decouples.

We consider in this paper the one-dimensional TASEP on a ring of $L$ sites. Each site is either empty or occupied by one classical particle. The dynamics consists of local hopping of the particles from one site $i$ to the next $i+1$ if the latter site is empty. Particles hop with rate $1$, \textit{i.e.} a particle has a probability $\rmd t$ to move in a small time interval $\rmd t$. The dynamics conserves the total number of particles $N$, and the average density $\overline{\rho}=N/L$ is constant in time. A configuration $\mathcal{C}$ of the system can be described by the occupation numbers of the sites $\eta_{i}\in\{0,1\}$, $i=1,\ldots,L$, where $\eta_{i}=1$ means that site $i$ is occupied and $\eta_{i}=0$ corresponds to an empty site. Equivalently, a configuration can be specified by the positions of the particles $x_{j}$, $j=1,\ldots,N$, $1\leq x_{1}<\ldots<x_{N}\leq L$.

The state of the system can also be described by a height function $H_{i}$, $i=1,\ldots,L$ in a mapping to an interface growth model. The mapping consists in evolving the initial height $H_{i}^{0}=\sum_{\ell=1}^{i}(\overline{\rho}-\eta_{\ell}^{0})$, built from the initial occupation numbers $\eta_{i}^{0}$ of TASEP, by the following dynamics: each time a particle moves from site $i$ to site $i+1$, $H_{i}$ increases by $1$. Extending the occupation numbers to a periodic function $\eta_{i}$, $i\in\mathbb{Z}$ of period $L$, the height $H_{i}$ is also periodic of period $L$ and verifies at all time $H_{i}=H_{i-1}+\overline{\rho}-\eta_{i}$ for any site $i$.

We are interested in the (total, time-integrated) current $Q$, equal to the number of times a particle has moved anywhere in the system between time $0$ and time $T$. This is a dynamical observable whose value can not be deduced from the knowledge of the positions of the particles in the system at time $T$ only, but depends also on the history from an initial state $\mathcal{C}_{0}$. It is however directly expressible from the height representation of TASEP as the difference between the final and the initial mean height, $Q/L=\tfrac{1}{L}\sum_{i=1}^{L}(H_{i}-H_{i}^{0})$.

The generating function of the current $\langle\rme^{\gamma Q}\rangle_{\mathcal{C}_{0}\to\mathcal{C}}$, where the averaging is taken over all realizations of the process conditioned on starting from initial configuration $\mathcal{C}_{0}$ at time $0$ and ending in configuration $\mathcal{C}$ at time $T$, obeys a (deformed) master equation \cite{DL1998.1}. In terms of the corresponding deformed Markov matrix $M(\gamma)$, the generating function is equal to \cite{P2015.1}
\begin{equation}
\label{G[M]}
\langle\rme^{\gamma Q}\rangle_{\mathcal{C}_{0}\to\mathcal{C}}=\frac{\langle\mathcal{C}|\rme^{TM(\gamma)}|\mathcal{C}_{0}\rangle}{\langle\mathcal{C}|\rme^{TM(0)}|\mathcal{C}_{0}\rangle}\;.
\end{equation}
The denominator, called $Z$ in the following, is the probability to observe the system in configuration $\mathcal{C}$ at time $T$ for an evolution starting in $\mathcal{C}_{0}$ at time $0$.

The problem is known to be integrable, as $M(\gamma)$ closely resembles the quantum Hamiltonian of the XXZ spin chain. It allows an exact treatment using Bethe ansatz to diagonalize $M(\gamma)$ and rewrite the generating function as a sum over normalized eigenstates
\begin{equation}
\label{G[E,psi]}
\langle\rme^{\gamma Q}\rangle_{\mathcal{C}_{0}\to\mathcal{C}}=\frac{1}{Z}\sum_{r}\rme^{TE_{r}(\gamma)}\langle\mathcal{C}|\psi_{r}\rangle\langle\psi_{r}|\mathcal{C}_{0}\rangle\;,
\end{equation}
with $Z$ ensuring that the generating function equals $1$ at $\gamma=0$. For finite systems, the eigenvalues and eigenvectors can be computed numerically very efficiently using exact Bethe ansatz formulas, which allows accurate evaluation of (\ref{G[E,psi]}) or other observables such as average density profile and current in a non-stationary setting \cite{MSS2012.1}, see also \cite{BPPP2006.1} for another approach based on an exact expression \cite{P2003.1} for the propagator of periodic ASEP. Bethe ansatz also allows exact calculations in the thermodynamic limit of large $L$, $N$ with fixed density $\overline{\rho}=N/L$, $0<\overline{\rho}<1$. This is especially true for periodic TASEP, for which the nearly decoupling structure of Bethe equations reduces enormously the complexity of the calculations. This has lead in the past to exact formulas for the spectral gap \cite{GS1992.1,GS1992.2,GM2004.1,GM2005.1} and large deviations of the current \cite{DL1998.1,DA1999.1}.

In order to study the thermodynamic limit of (\ref{G[E,psi]}), one needs to specify additionally how the final time and the initial and final configurations behave for large system size. The suitable scalings are known from KPZ universality \cite{SS2010.4,TSSS2011.1,KK2010.1,C2011.1,QS2015.1,HHT2015.1}, whose name comes from the Kardar-Parisi-Zhang equation \cite{KPZ1986.1}, and which describes universal features of the statistics of fluctuations in various interface growth models, driven-diffusive systems and directed polymers in random media. KPZ universality is characterized by spatial correlations on the scale $T^{2/3}$ for large time. We consider the relaxation scale $T\sim L^{3/2}$ on which the correlation length saturates to the full system size $L$. Initial and final conditions are then chosen to be well described by smooth density profiles on the full range of the system. This is however not sufficient due to propagation of density fluctuations around the system which hide the KPZ fluctuations generated by the dynamics that we are interested in. Over times $T\gg L$, these density fluctuations move ballistically at the velocity $1-2\,\overline{\rho}$. In order to correct for this, we take an initial configuration $\mathcal{C}_{0}$ corresponding to a fixed density profile $\rho_{\text{i}}$ and a final configuration $\mathcal{C}$ described by a density profile $\rho_{\text{f}}$ moving at velocity $1-2\,\overline{\rho}$. The current fluctuations are then defined by subtracting from $Q$ the deterministic part corresponding to the typical hydrodynamical evolution on the Euler time scale $T\sim L$ of the macroscopic density profile from Burgers' equation, described in section \ref{section Q deterministic} and \ref{appendix Burgers}.

With the previously mentioned scalings for the various quantities, the large $L$ limit of the summand of (\ref{G[E,psi]}) can be performed explicitly for the special cases of unit step \cite{P2015.2} and flat initial and final configurations, giving exact formulas for the generating function and probability density of current fluctuations. These exact results are extended to general step initial and final configurations with densities $\rho_{+}$ and $\rho_{-}$ by very accurate extrapolation of high precision finite size Bethe ansatz numerics. The main results are summarized in section \ref{section Q fluctuations}, with some technical details about Bethe ansatz relegated to \ref{appendix Bethe}.

Section \ref{section flat -> flat} is finally devoted to the special case of an evolution conditioned on flat initial and final states, for which the summation over eigenstates can be performed explicitly. It allows to extract the behaviour of current fluctuations $\xi_{t}$ when the rescaled time $t\propto T/L^{3/2}$ is small. With some proper definition of $t$ (\ref{scaling T}) and $\xi_{t}$ (\ref{xi t}), one finds the large deviations $P(\xi_{t}=t^{1/3}u)\sim\exp(-t^{-2/3}(C-\Xi(u)))$ with some known constant $C$, and $\Xi$ defined in (\ref{Xi[int,Ai]}). This is the main result of the paper. The rather technical saddle point analysis leading to it is carried out in \ref{appendix saddle point}.
\end{section}

\begin{section}{Deterministic leading orders of the current and Burgers' equation}
\label{section Q deterministic}
In this section, we summarize various known results about the deterministic evolution of the large scale density profile of TASEP on times $T\sim L$ from inviscid Burgers' equation. We deduce from this the deterministic leading orders for the total current on times $T\gg L$.

\begin{subsection}{Hydrodynamic evolution: inviscid Burgers' equation}
From the stochastic microscopic dynamics of TASEP, the occupation number $\eta_{i}$ of site $i$ evolves in time by
\begin{equation}
\label{detai/dT}
\frac{\rmd\langle\eta_{i}\rangle}{\rmd T}=\langle j_{i}\rangle-\langle j_{i+1}\rangle
\end{equation}
with an instantaneous current $j_{i}=\eta_{i-1}(1-\eta_{i})$. At large scales, a deterministic evolution emerges at leading order for the density profile $\rho(x,\tau)$, obtained by averaging occupation numbers $\eta_{i}$ over sites $i\simeq xL$. On the Euler time scale $T=\tau L$, the density profile evolves in time by a hyperbolic conservation law with one conserved quantity, the \textit{inviscid Burgers equation}
\begin{equation}
\label{Burgers}
\partial_{\tau}\rho+\partial_{x}j=0\;,
\end{equation}
with current-density relation
\begin{equation}
\label{j[rho]}
j=\rho(1-\rho)\;,
\end{equation}
and initial condition $\rho(x,0)=\rho_{0}(x)$ determined by the initial configuration of TASEP, see \textit{e.g.} \cite{S1991.1}. From time and space reversal in (\ref{detai/dT}), Burgers' equation also describes the macroscopic evolution for $\tau<0$ of TASEP conditioned on ending at time $T=0$ in a final configuration corresponding to a density profile $\rho_{1}$ of average $\overline{\rho}$: more precisely, the reversed profile $\tilde{\rho}(x,\tau)=\rho(1-x,-\tau)$ is the solution of Burgers' equation with initial condition $\tilde{\rho}(x,0)=\rho_{1}(1-x)$.

The solution to Burgers' equation (\ref{Burgers}) is only well defined locally in time, even with smooth initial condition: after a finite time, the solution $\rho(x,\tau)$ develops shocks, \textit{i.e.} discontinuities in $x$ at some point $z$ with a density lower on the left side of the shock $x<z$ than on the right side $x>z$. Indeed, the characteristics $x(\tau)$ such that $\rho(x(\tau),\tau)$ is constant in time (\textit{i.e.} $\rho(x(\tau),\tau)=\rho_{0}(x_{0})$ with $x_{0}=x(0)$) verify $x'(\tau)=1-2\rho_{0}(x_{0})$. In an interval $[x_{0},x_{1}]$ where $\rho_{0}$ decreases, it implies that the velocity of the characteristics starting at $x_{0}$ moves faster than the one starting at $x_{1}$, which leads to the formation of a discontinuity. This makes (\ref{Burgers}) ill-defined since the motion of the shock can not be derived from Burgers' equation. Unicity is recovered by imposing the additional constraint that the solution of (\ref{Burgers}) has to conserve the total density of particles $\overline{\rho}=\int_{0}^{1}\rmd x\,\rho(x,\tau)$ since the number of particles is conserved in TASEP. This is equivalent to considering the viscosity solution of (\ref{Burgers}), obtained by taking the limit of vanishing viscosity $\nu\to0$ in the solution of Burgers' equation with the additional viscosity term $\nu\partial_{x}^{2}\rho$ in the right hand side.
\end{subsection}

\begin{subsection}{Integrated current and height function}
On the Euler time scale $T=\tau L$, the total current per site up to time $T$ is equal at leading order in $L$ to $Q/L\simeq L\mathcal{Q}_{\tau}[\rho_{0}]$, with
\begin{equation}
\label{Q Burgers'}
\mathcal{Q}_{\tau}[\rho_{0}]=\int_{0}^{\tau}\rmd\sigma\,\int_{0}^{1}\rmd x\,j(x,\sigma)\;.
\end{equation}
The instantaneous current $j(x,\tau)$ is built from the current-density relation (\ref{j[rho]}) with $\rho(x,\tau)$ solution of (\ref{Burgers}) with initial condition $\rho_{0}$. Naively, the integral of $j(x,\tau)$ with respect to $x$ over the whole system is constant in time for the inviscid Burgers' equation since $\partial_{\tau}j(x,\tau)$ can be written as a derivative with respect to space as $\partial_{\tau}j=(1-2\rho)\partial_{\tau}\rho=-(1-2\rho)\partial_{x}j=\partial_{x}(1-2\rho)^{3}/6$. This argument breaks down after the formation of the first shock since then the integration over space has to be done between shocks whose positions depend on time.

As in the microscopic model, it is useful to define a height function associated to the density profile of the system by
\begin{equation}
\label{h}
h(x,\tau)=h_{0}(x)+\int_{0}^{\tau}\rmd\sigma\,j(x,\sigma)\;,
\end{equation}
with initial height equal to
\begin{equation}
\label{h0}
h_{0}(x)=\int_{0}^{x}\rmd y\,(\overline{\rho}-\rho_{0}(y))\;.
\end{equation}
This height function is equal to the large $L$ limit of the microscopic height of the interface of TASEP $H_{i}/L$ averaged over sites $i\simeq xL$. Burgers' integrated current (\ref{Q Burgers'}) is then related to the height function by
\begin{equation}
\label{Q[h]}
\mathcal{Q}_{\tau}=\overline{h}(\tau)-\overline{h}_{0}\;,
\end{equation}
with final and initial mean heights $\overline{h}(\tau)=\int_{0}^{1}\rmd x\,h(x,\tau)$ and $\overline{h}_{0}=\int_{0}^{1}\rmd x\,h_{0}(x)$.

From (\ref{Burgers}), the height function verifies $\partial_{\tau}h=\rho(1-\rho)$ and $\partial_{x}h=\overline{\rho}-\rho$, which implies that $h$ is solution of a deterministic KPZ equation without smoothing term $\partial_{\tau}h=\overline{\rho}(1-\overline{\rho})-(\partial_{x}h)^{2}-(1-2\,\overline{\rho})\partial_{x}h$.
\end{subsection}

\begin{subsection}{Large time evolution for smooth initial condition}
During the evolution, the number of shocks can increase when new shocks appear and decrease when consecutive shocks merge. With smooth initial density profile, the number of shocks become constant at some value $M\geq1$ at large time, generically $M=1$, the density profile converges to the flat profile of density $\overline{\rho}$, and the instantaneous current $j(x,\tau)$ converges for large $\tau$ to the stationary current
\begin{equation}
J=\overline{\rho}(1-\overline{\rho})\;.
\end{equation}
Burgers' total current (\ref{Q Burgers'}) is then approximatively equal to $\mathcal{Q}_{\tau}[\rho_{0}]\simeq J\tau$. We are interested in the corrections to this stationary value. They depend on the whole evolution of the density profile between time $0$ and time $\tau$, which involves in general the formation and merging of several shocks.

At large times, the density profile between consecutive shocks is approximatively given in the reference frame moving at the stationary speed of characteristics $1-2\,\overline{\rho}$ (called \textit{the moving frame} in the following) by ramps with negative slope of the form
\begin{equation}
\label{rho(x,tau) large tau}
\rho(x+(1-2\,\overline{\rho})\tau,\tau)\simeq\overline{\rho}-\frac{x-\kappa}{2\tau}\;.
\end{equation}
The position $\kappa$, which corresponds to a density exactly equal to $\overline{\rho}$, is located somewhere between the two shocks considered.

In the generic case where only one shock remains at large enough time, its position is equal to $z(\tau)\simeq(1-2\,\overline{\rho})\tau+\kappa+\half$ modulo $1$ by conservation of the density. If $M\geq2$ shocks survive at large times and never merge afterwards, their positions in the moving frame are equal to $(\kappa_{-}+\kappa_{+})/2$ with $\kappa_{-}$ and $\kappa_{+}$ the positions at which the ramps on the left and on the right side of the shock have density $\overline{\rho}$.

The large time behaviour of the system is thus governed by the positions $\kappa$ at which the ramps have density $\overline{\rho}$. Each number $\kappa$ is the initial point of a characteristics of the partial differential equation (\ref{Burgers}) that never meets shocks and thus exists for all times. Such characteristics are called \textit{divides} \cite{D2010.3}. They have been defined more generally for hyperbolic conservation laws with concave (or convex) current-density relation, of which Burgers' equation (\ref{Burgers}) is the simplest non-trivial example. The initial points $\kappa$ of divides are the solutions of $\rho_{0}(\kappa)=\overline{\rho}$ such that $\int_{\kappa}^{x}\rmd y\,(\rho_{0}(y)-\overline{\rho})\leq0$ for all $x$, see \cite{D2010.3} theorem 11.4.1. Equivalently, they are the locations of the global minima of the initial height profile $h_{0}$ defined in (\ref{h0}):
\begin{equation}
\label{kappa}
h_{0}(\kappa)=\min_{x}h_{0}(x)\;.
\end{equation}
This is physically reasonable for TASEP since in the mapping to an interface growth model, the height only grows from local minima of the interface, and thus one expects that the large time behaviour is governed by the global minima of the initial interface.

In the generic case where only one shock subsists at large times, $\kappa$ is unique and is equal to the position in the moving frame of the center of the ramp to which the density profile converges. For non-generic initial profiles, $h_{0}$ can have $M\geq2$ global minima, which leads at large time to the existence of $M$ shocks. The special case of a flat initial density profile $\rho_{0}(x)=\overline{\rho}$ corresponds to a situation with no shocks.
\end{subsection}

\begin{subsection}{Burgers' current at large time}
From the relation $\partial_{x}h=\overline{\rho}-\rho$, the height $h(x,\tau)$ can be written as an integral over space with upper bound $x$. The constant of integration is obtained from $h((1-2\,\overline{\rho})\tau+\kappa,\tau)=J\tau+h_{0}(\kappa)$, which follows from taking the derivative with respect to $\tau$ of $h((1-2\,\overline{\rho})\tau+\kappa,\tau)$ and using the fact that characteristics starting from $\kappa$ have density $\overline{\rho}$. One finds the expansion
\begin{equation}
\label{h large tau}
h(x+(1-2\,\overline{\rho})\tau,\tau)\simeq
J\tau+h_{0}(\kappa)+\frac{(x-\kappa)^{2}}{4\tau}\;,
\end{equation}
for $x$ inside the interval between two consecutive shocks corresponding to the ramp associated to a global minimum $\kappa$ of $h_{0}$. Integrating with respect to $x$ for each ramp, (\ref{Q[h]}) gives an expansion for Burgers' current $\mathcal{Q}_{\tau}$. In the generic case where the global minimum of $h_{0}$ is unique, one finds
\begin{equation}
\label{Q Burgers' inviscid asymptotics}
\mathcal{Q}_{\tau}[\rho_{0}]\simeq J\tau+\mathcal{R}[\rho_{0}]+\frac{1}{48\tau}\;,
\end{equation}
with
\begin{equation}
\label{R}
\mathcal{R}[\rho_{0}]=\min_{x}h_{0}(x)-\overline{h}_{0}=-\int_{0}^{1}\rmd x\,x(\rho_{0}(\kappa+x)-\overline{\rho})\;.
\end{equation}
The quantity $\mathcal{R}[\rho_{0}]$ vanishes for a flat initial profile $\rho_{0}(x)=\overline{\rho}$. Furthermore, for any initial profile, one has $\mathcal{R}[\rho_{0}]\leq0$: the particles in TASEP move less easily on average when the density profile is not flat, which reduces the total integrated current. The expression (\ref{R}) is checked in \ref{appendix Burgers} for some simple piecewise linear initial density profiles by solving explicitly Burgers' equation and calculating the current at finite $\tau$ from (\ref{Q Burgers'}).

If $h_{0}$ has $M\geq2$ global minima, the term of order $1/\tau$ in (\ref{Q Burgers' inviscid asymptotics}) is replaced by $\sum_{k=1}^{M}\frac{\lambda_{k}^{3}}{6\tau}$ with $\lambda_{k}$ the length of the interval for which the $k$-th ramp has density larger than $\overline{\rho}$. With flat initial condition $\rho_{0}(x)=\overline{\rho}$, the current is exactly equal to $J\tau$ with no higher order correction.
\end{subsection}

\begin{subsection}{Deterministic current for TASEP conditioned on the initial and the final state}
Burgers' equation describes the deterministic current for TASEP on a time scale $T\sim L$. On a longer time scale $T\gg L$, the macroscopic density profile stays flat for essentially all the evolution, leading to a total current per site equal to $Q/L=JT$ at leading order in $L$. Considering an evolution conditioned to start at time $0$ in a configuration corresponding to a fixed density profile $\rho_{\text{i}}$ and to end at time $T$ in a configuration corresponding to a density profile $\rho_{\text{f}}$ in the moving frame, the first correction to the stationary value of the current comes from time intervals with size of order $L$ at the beginning and the end of the evolution. It is expressed in terms of the quantity $\mathcal{R}$ defined in (\ref{R}) as
\begin{equation}
\label{Qdet}
\frac{Q}{L}\simeq\frac{Q_{\text{det}}}{L}=JT+(\mathcal{R}[\rho_{\text{i}}]+\mathcal{R}[\tilde{\rho}_{\text{f}}])L\;,
\end{equation}
with $\tilde{\rho}_{\text{f}}(x)=\rho_{\text{f}}(1-x)$.

In the next section, we study the fluctuations of $Q$ beyond the deterministic value (\ref{Qdet}) on the KPZ time scale $T\sim L^{3/2}$ using results from Bethe ansatz for specific initial and final states.
\end{subsection}
\end{section}

\begin{section}{Fluctuations}
\label{section Q fluctuations}
On the KPZ time scale $T\sim L^{3/2}$, the density profile is typically equal to the constant profile $\overline{\rho}$, except for small time intervals of duration $\sim L$ at the beginning and at the end of the time range, where the density profile evolves from Burgers' equation (\ref{Burgers}). From KPZ universality, height fluctuations in the moving frame have an amplitude $T^{1/3}\sim\sqrt{L}$ and are correlated on the spatial scale $T^{2/3}\sim L$. We define the rescaled time
\begin{equation}
\label{scaling T}
T=\frac{t\,L^{3/2}}{\sqrt{\overline{\rho}(1-\overline{\rho})}}\;.
\end{equation}

\begin{subsection}{Generating function}
We consider a fixed initial configuration $\mathcal{C}_{0}$ corresponding at large scale to the density profile $\rho_{\text{i}}$, and a final configuration $\mathcal{C}$ corresponding in the moving frame to the density profile $\rho_{\text{f}}$ independent of $T$. All density profiles are periodic with periodicity $1$. Based on the results of section \ref{section Q deterministic} and on the scaling of height fluctuation in KPZ universality, we define current fluctuations as
\begin{equation}
\label{xi t}
\xi_{t}=\frac{Q-Q_{\text{det}}}{\sqrt{\overline{\rho}(1-\overline{\rho})}L^{3/2}}\;,
\end{equation}
with the deterministic value of the current $Q_{\text{det}}$ given by (\ref{Qdet}).

We are interested in the statistics of the random variable $\xi_{t}$. We consider the generating function (\ref{G[M]}), (\ref{G[E,psi]}) with fugacity
\begin{equation}
\label{scaling gamma}
\gamma=\frac{s}{\sqrt{\overline{\rho}(1-\overline{\rho})}L^{3/2}}\;.
\end{equation}
From KPZ universality, one expects that
\begin{equation}
\langle\rme^{\gamma(Q-Q_{\text{det}})}\rangle_{\mathcal{C}_{0}\to\mathcal{C}}=\langle\rme^{s\xi_{t}}\rangle_{\mathcal{C}_{0}\to\mathcal{C}}\;
\end{equation}
has a finite limit when $L\to\infty$ with the scaling (\ref{scaling T}) for $T$, and initial and final configurations corresponding to fixed density profiles in the reference frames described above. We define
\begin{equation}
G_{t}(s)=\lim_{L\to\infty}\langle\rme^{s\xi_{t}}\rangle_{\mathcal{C}_{0}\to\mathcal{C}}\;.
\end{equation}
The average over histories in the previous equation can be computed from the decomposition (\ref{G[E,psi]}) over normalized eigenstates. One has
\begin{equation}
\label{G[E(gamma),psi(gamma)]}
G_{t}(s)=\lim_{L\to\infty}\frac{\rme^{-\gamma Q_{\text{det}}}}{Z}\sum_{r}\rme^{TE_{r}(\gamma)}\langle\mathcal{C}|\psi_{r}(\gamma)\rangle\langle\psi_{r}(\gamma)|\mathcal{C}_{0}\rangle\;.
\end{equation}
with
\begin{equation}
\label{Z[E(0),psi(0)]}
Z=\sum_{r}\rme^{TE_{r}(0)}\langle\mathcal{C}|\psi_{r}(0)\rangle\langle\psi_{r}(0)|\mathcal{C}_{0}\rangle\;
\end{equation}
the probability to find the system in configuration $\mathcal{C}$ at time $T$ for an initial configuration $\mathcal{C}_{0}$. Typical eigenvalues $E$ of the Markov matrix $M(0)$ scale as $E\sim L$ with $\Re(E/L)<0$, see figure \ref{Fig spectrum}. Since the number of eigenvalues with a given value of $E/L$ is of order $\exp(\mathfrak{s}L)$ \cite{P2013.1} with finite "entropy" $\mathfrak{s}=\mathfrak{s}(E/L)$, these typical eigenvalues have a vanishing contribution to $Z$. Extrapolating the small $E/L$ behaviour $\mathfrak{s}(E/L)\sim|E/L|^{2/5}$ \cite{P2013.1} to eigenvalues $E\sim L^{\alpha}$, $\alpha<1$ closer to the stationary eigenvalue $0$, we observe that if $\alpha>-3/2$, the contribution of the entropy $\exp(\tilde{\mathfrak{s}}L^{(3+2\alpha)/5})$, $\tilde{\mathfrak{s}}>0$ can not compensate the vanishingly small contribution of $TE_{r}$, equal to $\exp(\tilde{\mathfrak{e}}L^{\alpha+3/2})$, $\Re\,\tilde{\mathfrak{e}}<0$. Furthermore, the eigenvalues with largest non-zero real part scale as $L^{-3/2}$ \cite{GS1992.1}. Therefore, only the eigenstates whose eigenvalues have a real part scaling as $L^{-3/2}$ contribute to (\ref{Z[E(0),psi(0)]}). These eigenvalues correspond to the tip of the peak located at $0$ in figure \ref{Fig spectrum}. The same kind of reasoning can presumably be used for non-zero $\gamma\sim L^{-3/2}$ too, for which the scalings for the entropy of eigenvalues should not be modified.
\begin{figure}
  \begin{center}
    \begin{tabular}{lll}
      \begin{tabular}{c}\includegraphics[width=75mm]{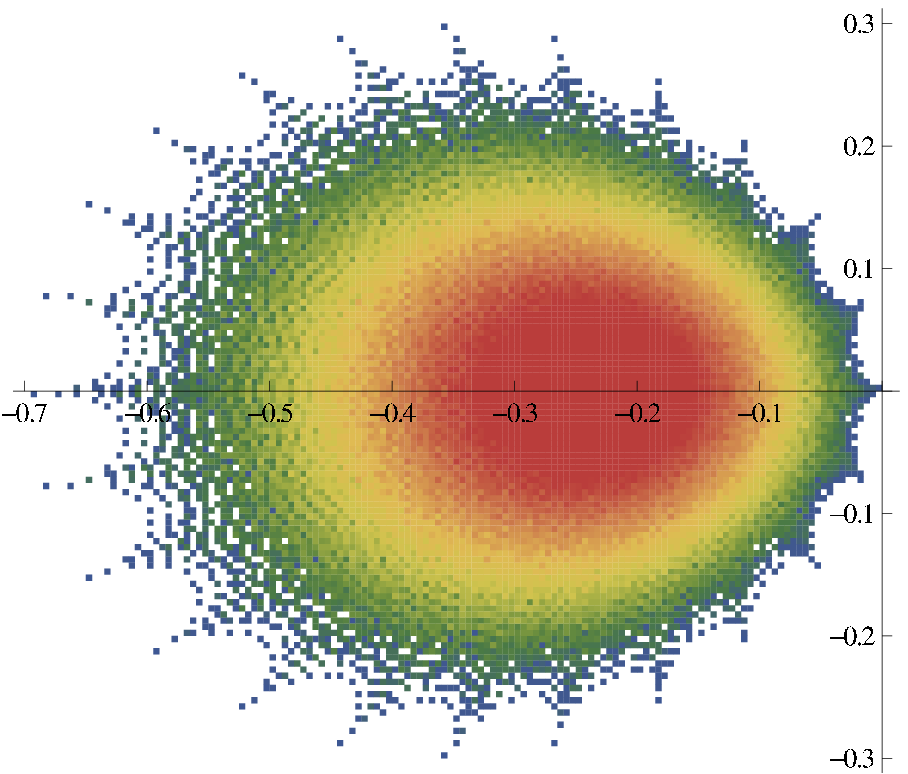}\end{tabular}
      &&
      \begin{tabular}{c}\includegraphics[width=60mm]{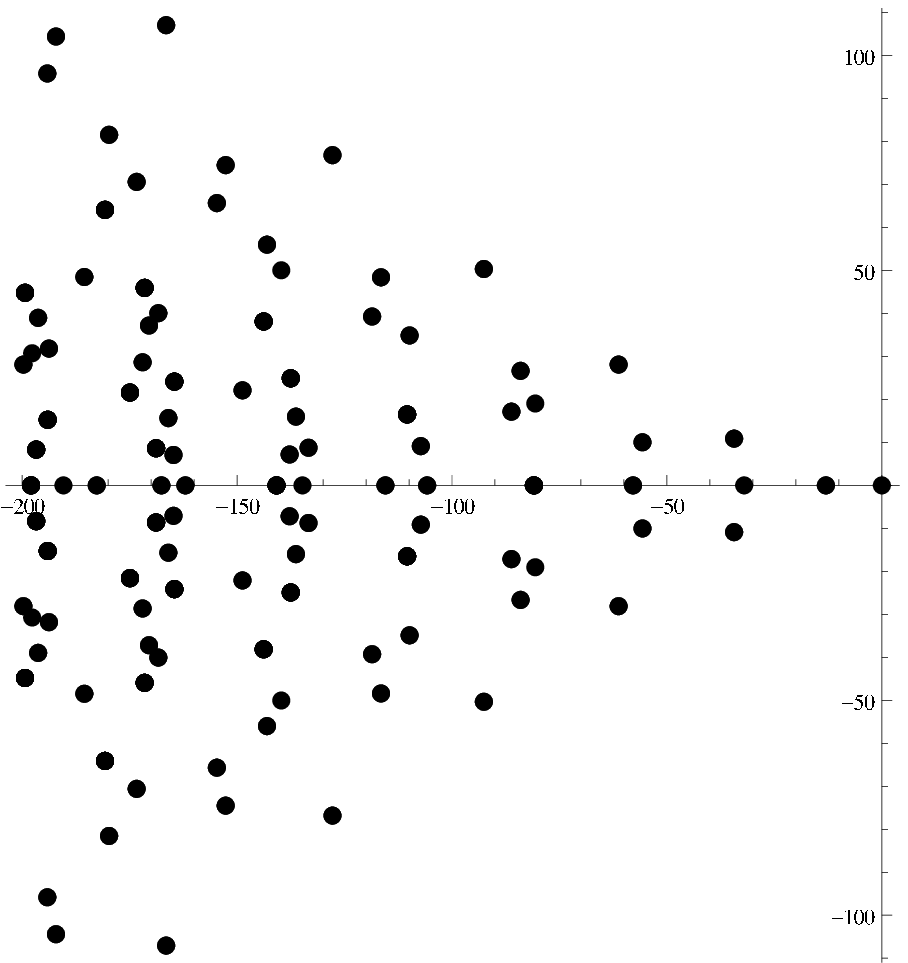}\end{tabular}
    \end{tabular}
  \end{center}
  \caption{Spectrum of TASEP at half-filling $\overline\rho=1/2$. The graph on the left represents in the complex plane the eigenvalues (divided by $L$) of the Markov matrix $M(0)$ with $N=12$ particles on $L=24$ sites. Brighter colors in the middle correspond to many eigenvalues, darker colors on the borders to fewer eigenvalues. The graph on the right corresponds to the asymptotics (\ref{E asymptotics}) $\chi_{r}(2\pi c_{r})$ at rescaled fugacity $s=0$ of the eigenvalues closest to $0$.}
  \label{Fig spectrum}
\end{figure}

In the following, we call \textit{first eigenstates} the infinitely many eigenstates whose eigenvalue has a real part scaling as $L^{-3/2}$ when setting $s=0$ (for $s\neq0$, the real part of the eigenvalues gains a term of order $L^{-1/2}$, see (\ref{E asymptotics}), but this term does not depend on the eigenstate and thus factors out of (\ref{G[E(gamma),psi(gamma)]})). From Bethe ansatz, each eigenstate is characterized by $N$ pseudo-momenta $k_{j}$, $j=1,\ldots,N$, integers or half-integers depending on the parity of $N$. For the stationary state, the pseudo-momenta form a Fermi sea, $k_{j}^{0}=j-(N+1)/2$. The first eigenstates can be understood as particle/hole excitations over this Fermi sea \cite{P2014.1}, corresponding to moving some pseudo-momenta with $|k_{j}^{0}|<N/2$ close to $\pm N/2$ to excited values with $|k_{j}|>N/2$, still close to $\pm N/2$. These excitations can be conveniently labelled by $4$ finite sets of positive half-integers $A_{0}^{\pm}$ and $A^{\pm}$ representing respectively the positions of the hole and particle excitations on both sides of the Fermi sea, see figure \ref{Fig sets A}. Each creation of a hole on one side of the Fermi sea must be accompanied by the creation of a particle on the same side of the Fermi sea for the first eigenstates: any imbalance leads to eigenvalues with real part scaling as $L^{\alpha}$ with some $\alpha>-3/2$. It implies that the cardinals of the sets verify the constraints
\begin{equation}
\label{m+-}
m_{r}^{+}\equiv|A_{0}^{+}|=|A^{+}|
\qquad\text{and}\qquad
m_{r}^{-}\equiv|A_{0}^{-}|=|A^{-}|\;.
\end{equation}
In the following, the four sets $A_{0}^{\pm}$, $A^{\pm}$ are collectively denoted by the index $r$.
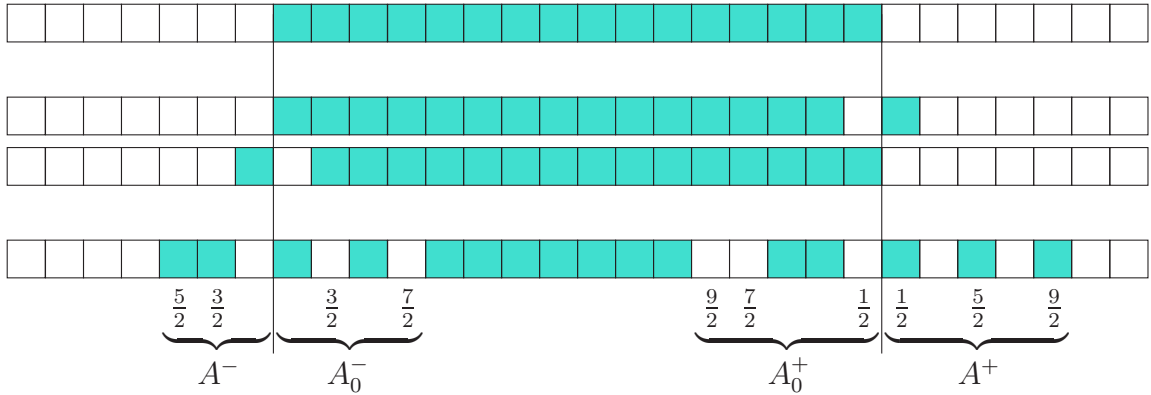
\begin{figure}
  \begin{center}
    \begin{tabular}{c}
      \begin{picture}(150,5)
        \put(35,0){\color[rgb]{0.25,0.875,0.8125}\polygon*(0,0)(80,0)(80,5)(0,5)}
        \multiput(0,0)(5,0){30}{\polygon(0,0)(5,0)(5,5)(0,5)}
      \end{picture}\\\\
      \begin{picture}(150,5)
        \put(35,0){\color[rgb]{0.25,0.875,0.8125}\polygon*(0,0)(75,0)(75,5)(0,5)}
        \put(115,0){\color[rgb]{0.25,0.875,0.8125}\polygon*(0,0)(5,0)(5,5)(0,5)}
        \multiput(0,0)(5,0){30}{\polygon(0,0)(5,0)(5,5)(0,5)}
      \end{picture}\\
      \begin{picture}(150,5)
        \put(40,0){\color[rgb]{0.25,0.875,0.8125}\polygon*(0,0)(75,0)(75,5)(0,5)}
        \put(30,0){\color[rgb]{0.25,0.875,0.8125}\polygon*(0,0)(5,0)(5,5)(0,5)}
        \multiput(0,0)(5,0){30}{\polygon(0,0)(5,0)(5,5)(0,5)}
      \end{picture}\\\\
      \begin{picture}(150,5)
        \put(55,0){\color[rgb]{0.25,0.875,0.8125}\polygon*(0,0)(35,0)(35,5)(0,5)}
        \put(20,0){\color[rgb]{0.25,0.875,0.8125}\polygon*(0,0)(10,0)(10,5)(0,5)}
        \put(35,0){\color[rgb]{0.25,0.875,0.8125}\polygon*(0,0)(5,0)(5,5)(0,5)}
        \put(45,0){\color[rgb]{0.25,0.875,0.8125}\polygon*(0,0)(5,0)(5,5)(0,5)}
        \put(100,0){\color[rgb]{0.25,0.875,0.8125}\polygon*(0,0)(10,0)(10,5)(0,5)}
        \put(115,0){\color[rgb]{0.25,0.875,0.8125}\polygon*(0,0)(5,0)(5,5)(0,5)}
        \put(125,0){\color[rgb]{0.25,0.875,0.8125}\polygon*(0,0)(5,0)(5,5)(0,5)}
        \put(135,0){\color[rgb]{0.25,0.875,0.8125}\polygon*(0,0)(5,0)(5,5)(0,5)}
        \multiput(0,0)(5,0){30}{\polygon(0,0)(5,0)(5,5)(0,5)}
        \put(21.5,-5){$\frac{5}{2}$}
        \put(26.5,-5){$\frac{3}{2}$}
        \put(41.5,-5){$\frac{3}{2}$}
        \put(51.5,-5){$\frac{7}{2}$}
        \put(91.5,-5){$\frac{9}{2}$}
        \put(96.5,-5){$\frac{7}{2}$}
        \put(111.5,-5){$\frac{1}{2}$}
        \put(116.5,-5){$\frac{1}{2}$}
        \put(126.5,-5){$\frac{5}{2}$}
        \put(136.5,-5){$\frac{9}{2}$}
        \put(20.5,-7){$\underbrace{\hspace{14\unitlength}}$}
        \put(35.5,-7){$\underbrace{\hspace{19\unitlength}}$}
        \put(90.5,-7){$\underbrace{\hspace{24\unitlength}}$}
        \put(115.5,-7){$\underbrace{\hspace{24\unitlength}}$}
        \put(25,-14){$A^{-}$}
        \put(42,-14){$A_{0}^{-}$}
        \put(100,-14){$A_{0}^{+}$}
        \put(125,-14){$A^{+}$}
        \put(35,-10){\line(0,1){45}}
        \put(115,-10){\line(0,1){45}}
      \end{picture}\\\\
      \hspace*{0mm}
    \end{tabular}
  \end{center}
  \caption{Graphical representation of the (half-)integers $k_{j}$ (coloured squares) characterizing some of the \textit{first eigenstates}, whose eigenvalue has a real part scaling as $L^{-3/2}$. From top to bottom, the pictures correspond to the stationary state, the two eigenstates giving the spectral gap, and a generic eigenstate with particle-hole excitations described by four sets of half-integers $A_{0}^{\pm}$, $A^{\pm}$.}
  \label{Fig sets A}
\end{figure}
\end{subsection}

\begin{subsection}{Large $L$ asymptotics}
For each first eigenstate $r$, it is convenient to introduce the function $\chi_{r}$, with branch cuts $\rmi[\pi,\infty)$ and $-\rmi[\pi,\infty)$, defined by
\begin{eqnarray}
\label{chi[A,zeta]}
&& \chi_{r}(u)=\frac{8\pi^{3/2}}{3}\Big(\rme^{-\rmi\pi/4}\zeta\big(-\frac{3}{2},\frac{1}{2}+\frac{\rmi u}{2\pi}\big)+\rme^{\rmi\pi/4}\zeta\big(-\frac{3}{2},\frac{1}{2}-\frac{\rmi u}{2\pi}\big)\Big)\nonumber\\
&&\hspace{13mm} -\frac{8\pi^{3/2}}{3}\Bigg(\sum_{a\in A_{0}^{+}}\sqrt{\rmi}\,\Big(a-\frac{\rmi u}{2\pi}\Big)^{3/2}+\sum_{a\in A^{-}}\sqrt{\rmi}\,\Big(a-\frac{\rmi u}{2\pi}\Big)^{3/2}\\
&&\hspace{32mm} +\sum_{a\in A_{0}^{-}}\sqrt{-\rmi}\,\Big(a+\frac{\rmi u}{2\pi}\Big)^{3/2}+\sum_{a\in A^{+}}\sqrt{-\rmi}\,\Big(a+\frac{\rmi u}{2\pi}\Big)^{3/2}\Bigg)\;,\nonumber
\end{eqnarray}
where $\zeta$ is the Hurwitz zeta function. For the stationary state, the four sets are empty, and the function reduces to a polylogarithm from Jonqui\`ere's identity: $\chi_{0}(u)=-(2\pi)^{-1/2}\Li_{5/2}(-\rme^{u})$ if $-\pi<\Im\,u<\pi$. We also introduce the complex number $c_{r}\equiv c_{r}(s)$, solution of
\begin{equation}
\label{c(s)}
\chi_{r}'(2\pi c_{r})=s\;.
\end{equation}
Neither existence nor unicity of $c_{r}$ has been proved; numerics seem however to indicate that both hold for any choice of the sets satisfying (\ref{m+-}) if the rescaled fugacity $s$ verifies $\Re\,s\geq0$, which is a consequence of the restriction $\Re\,\gamma\geq0$ of \ref{appendix Bethe}. The stationary state has the singular solution $c_{0}(s)\to-\infty$ when $s\to0$. For some choices of the sets $A_{0}^{\pm}$, $A^{\pm}$, the solution $c_{r}$ can have very large imaginary part, which makes an analytic continuation needed if one wants to work with polylogarithms instead of Hurwitz $\zeta$ functions.

The eigenvalues of the first eigenstates have the large $L$ expansion \cite{P2014.1}
\begin{equation}
\label{E asymptotics}
E_{r}(\gamma)\simeq\frac{s\sqrt{\overline{\rho}(1-\overline{\rho})}}{\sqrt{L}}-\frac{2\rmi\pi(1-2\,\overline{\rho})p_{r}}{L}+\frac{\sqrt{\overline{\rho}(1-\overline{\rho})}}{L^{3/2}}\,\chi_{r}(2\pi c_{r})\;,
\end{equation}
see figure \ref{Fig spectrum} for a graphical representation of the first few $\chi_{r}(2\pi c_{r})$. The special case of the spectral gap, corresponding to the first non-zero eigenvalue, was obtained in \cite{GS1992.1,GS1992.2,GM2004.1,GM2005.1}. It has been also studied for periodic ASEP \cite{K1995.1}, for TASEP \cite{dGE2005.1,dGE2006.1} and ASEP \cite{dGE2008.1,dGFS2011.1} on an open interval, and for periodic ASEP with several species of particles \cite{AKSS2009.1,WK2010.1}.

We consider unnormalized Bethe eigenvectors, described more precisely in \ref{appendix Bethe}. The left and right eigenvectors $\phi_{r}$ can be chosen in such a way that
\begin{equation}
\label{phiL[phiR]}
\langle\vec{x}|\phi_{r}\rangle=\langle\phi_{r}|\tilde{\vec{x}}\rangle\;,
\end{equation}
where the configuration $\vec{x}$ with particle at positions $1\leq x_{1}<\ldots<x_{N}\leq L$ and the configuration $\tilde{\vec{x}}$ with particle at positions $1\leq\tilde{x}_{1}<\ldots<\tilde{x}_{N}\leq L$ are related by space reversal $\tilde{x}_{j}=L+1-x_{N+1-j}$. The large $L$ limit for the normalization of these Bethe eigenstates has been obtained in \cite{P2015.2}:
\begin{equation}
\label{norm phi}
\frac{\Omega}{\langle\phi_{r}|\phi_{r}\rangle}\simeq\frac{\rme^{2\pi c_{r}}}{\sqrt{2\pi}\,\chi_{r}''(2\pi c_{r})}\;,
\end{equation}
with $\Omega=\C{L}{N}$ the total number of configurations. We changed the overall normalization of the eigenstates from \cite{P2015.2} in order to make the elements of the eigenvectors simpler, see \ref{appendix Bethe}.

For a configuration $\mathcal{C}$ corresponding to a fixed density profile $\rho_{0}$, we write the asymptotics of the elements of the eigenvectors as
\begin{equation}
\label{phi asymptotics[R,Phi]}
\langle\phi_{r}|\mathcal{C}\rangle\simeq \rme^{\mathcal{R}[\rho_{0}]\gamma L^{2}}\,\Phi_{r}[\rho_{0}]\;.
\end{equation}
Shifting a configuration by a distance $X$ gives an additional factor $\rme^{2\rmi\pi p_{r}X/L}$ to $\langle\mathcal{C}|\phi_{r}\rangle$, with in particular $X=(1-2\,\overline{\rho})T$ for a configuration corresponding to a density profile fixed in the moving frame. Gathering everything, this implies for the generating function of current fluctuations
\begin{equation}
\label{G[c,Phi]}
G_{t}(s)=\frac{1}{\mathcal{Z}_{t}}\sum_{r}\frac{\rme^{2\pi c_{r}}\rme^{t\chi_{r}(2\pi c_{r})}}{\sqrt{2\pi}\,\chi_{r}''(2\pi c_{r})}\,\Phi_{r}[\tilde{\rho}_{\text{f}}]\,\Phi_{r}[\rho_{\text{i}}]\;,
\end{equation}
with normalization constant $\mathcal{Z}_{t}=\Omega Z$ equal to the probability of having the system in configuration $\mathcal{C}$ at time $T$ starting in $\mathcal{C}_{0}$, divided by the stationary probability $\Omega^{-1}$. The stationary eigenvector $r=0$ at fugacity $\gamma=0$ verifies $\Phi_{0}[\rho]=1$ independently of $\rho$, and $\mathcal{Z}_{t}\to1$ when $t\to\infty$ since $c_{0}(s)\to-\infty$ when $s\to0$.

From section \ref{section Q deterministic}, the quantity $\Phi_{r}[\rho_{0}]$ is expected to be independent of $L$ and to depend only on $\rho_{0}$ and not on the details of the configuration $\mathcal{C}$. It can be computed explicitly in the special cases where the Bethe ansatz expressions for the eigenvectors reduce to Vandermonde determinants. This is in particular the case for flat and unit step configurations, which leads to exact formulas for the current fluctuations in four cases, denoted flat $\to$ flat, step $\to$ flat, flat $\to$ step, step $\to$ step, depending on the initial and the final state on which the evolution is conditioned. The flat $\to$ flat case is studied in much detail in section \ref{section flat -> flat}.

\begin{subsubsection}{Flat configurations}
The component of the eigenvector is computed by elementary manipulations in \ref{appendix Bethe} for a flat configuration $\mathcal{F}$ with particles at positions $x_{j}=X+(j-1)/\overline{\rho}$, $j=1,\ldots,N$ and $\overline{\rho}^{-1}$ integer, which corresponds at large scale to a flat profile $\rho^{\mathcal{F}}(x)=\overline{\rho}$. One has $\mathcal{R}[\rho^{\mathcal{F}}]=0$ and
\begin{equation}
\Phi_{r}[\rho^{\mathcal{F}}]=\Phi_{r}^{\mathcal{F}}\;
\end{equation}
independently of $X$, with
\begin{equation}
\label{PhiF}
\Phi_{r}^{\mathcal{F}}
=\openone_{\{A_{0}^{+}=A^{-}\}}
\openone_{\{A_{0}^{-}=A^{+}\}}
\,\frac{\rmi^{m_{r}}}{(1+\rme^{2\pi c_{r}})^{1/4}}\;.
\end{equation}
The constraint $A_{0}^{+}=A^{-}$, $A_{0}^{-}=A^{+}$ implies $m_{r}=m_{r}^{+}+m_{r}^{-}$. The elements of the eigenvector corresponding to flat configurations vanish exactly when $A_{0}^{+}\neq A^{-}$ or $A_{0}^{-}\neq A^{+}$ because of the symmetries of the configuration.

The factor $(1+\rme^{2\pi c})^{-1/4}$ in (\ref{PhiF}) is understood with the same branch cuts $\rmi[\half,\infty)$ and $-\rmi[\half,\infty)$ as $\chi_{r}(2\pi c)$. With the usual definition of the non-integer power $z^{1/4}=\exp(\tfrac{1}{4}\log z)$ and the usual branch cut $\mathbb{R}^{-}$ for the logarithm, the factor $(1+\rme^{2\pi c})^{-1/4}$ is interpreted as $(-\rmi)^{\lfloor\Im(c+\rmi/2)\rfloor}(1+\rme^{2\pi c})^{-1/4}$ with $\lfloor x\rfloor$ the largest integer lower than $x$.

The expression (\ref{PhiF}) has been checked numerically using rational Richardson extrapolation \cite{HS1988.1} (also called the Bulirsch-Stoer method) of finite size Bethe ansatz numerics. Richardson extrapolation allows to extract the constant term $f_{0}$ of an expansion of the form $f(L)=\sum_{k=0}^{\infty}f_{k}L^{-k\,\omega}$ knowing a few values $f(L)$ (with high precision) for moderate values of $L$. It often allows to extract around $10$ correct digits of $f(0)$ knowing $20$ values $f(L)$ with only one significant digit in common with $f_{0}$, see table \ref{Table extrapolation} for an example. Richardson extrapolation comes naturally with an accurate estimator for the error on $f_{0}$. It was used here not only for the configuration $\mathcal{F}$, but also for more general configurations corresponding to a macroscopic flat profile, built by repeating clusters of the form $1^{r_{+}}0^{r_{-}}$ with $r_{+},r_{-}>0$, $r_{+}+r_{-}\leq5$. A perfect agreement was found with (\ref{PhiF}) within at least $10$ digits, see table \ref{Table extrapolation} for an example. All the computations were done with a generic value $s=0.2+\rmi$ for the rescaled asymmetry. The natural exponent $\omega=1/2$ was used for the extrapolation.
\begin{table}
  \begin{center}
  \begin{tabular}{lllll}
    $L$ && Numerical value && Richardson extrapolation\\
    4 && $0.646361\, -0.409949\,\rmi$ && $0.-0.\rmi$ \\
    8 && $0.658283\, -0.403781\,\rmi$ && $0.7-0.4\,\rmi$ \\
    12 && $0.66433\, -0.398974\,\rmi$ && $0.7-0.3\,\rmi$ \\
    16 && $0.668107\, -0.39556\,\rmi$ && $0.7-0.4\,\rmi$ \\
    20 && $0.670745\, -0.393016\,\rmi$ && $0.69-0.37\,\rmi$ \\
    24 && $0.67272\, -0.391037\,\rmi$ && $0.694-0.367\,\rmi$ \\
    28 && $0.674269\, -0.389443\,\rmi$ && $0.6941-0.3667\,\rmi$ \\
    32 && $0.675525\, -0.388125\,\rmi$ && $0.6940-0.3666\,\rmi$ \\
    36 && $0.67657\, -0.387012\,\rmi$ && $0.6941-0.3667\,\rmi$ \\
    40 && $0.677457\, -0.386056\,\rmi$ && $0.69407-0.36669\,\rmi$ \\
    44 && $0.678222\, -0.385223\,\rmi$ && $0.694065-0.366690\,\rmi$ \\
    48 && $0.678891\, -0.384489\,\rmi$ && $0.6940651-0.3666903\,\rmi$ \\
    52 && $0.679482\, -0.383836\,\rmi$ && $0.6940651-0.3666903\,\rmi$ \\
    56 && $0.68001\, -0.383251\,\rmi$ && $0.6940651-0.3666903\,\rmi$ \\
    60 && $0.680484\, -0.382721\,\rmi$ && $0.694065124-0.366690292\,\rmi$ \\
    64 && $0.680914\, -0.382239\,\rmi$ && $0.694065123-0.366690292\,\rmi$ \\
    68 && $0.681305\, -0.381799\,\rmi$ && $0.6940651235-0.3666902918\,\rmi$ \\
    72 && $0.681663\, -0.381394\,\rmi$ && $0.6940651235-0.3666902918\,\rmi$ \\
    76 && $0.681993\, -0.38102\,\rmi$ && $0.69406512350-0.36669029178\,\rmi$ \\
    80 && $0.682298\, -0.380673\,\rmi$ && $0.694065123499-0.366690291776\,\rmi$ \\
    84 && $0.682582\, -0.38035\,\rmi$ && $0.6940651234993-0.3666902917760\,\rmi$ \\
    88 && $0.682845\, -0.380049\,\rmi$ && $0.6940651234993-0.3666902917760\,\rmi$ \\
    92 && $0.683092\, -0.379766\,\rmi$ && $0.6940651234993-0.3666902917760\,\rmi$ \\
    96 && $0.683323\, -0.379501\,\rmi$ && $0.694065123499272-0.366690291775979\,\rmi$ \\
    100 && $0.68354\, -0.379252\,\rmi$ && $0.694065123499272-0.366690291775980\,\rmi$ \\
  \end{tabular}
  \end{center}
  \caption{Richardson extrapolation of finite size Bethe ansatz numerics for the component of the right stationary eigenvector corresponding to a configuration $\mathcal{C}$ of the form $\bullet\bullet\textunderscore\,\textunderscore\bullet\bullet\textunderscore\,\textunderscore\ldots\bullet\bullet\textunderscore\,\textunderscore$, where $\bullet$ corresponds to a particle and $\textunderscore$ to an empty site. The calculations are done with $50$ digit precision and the rescaled fugacity is equal to the generic value $s=0.2+\rmi$. The first column corresponds to the size $L$ of the system and the second column to (the first digits of) the numerical evaluation of $\langle\mathcal{C}|\phi_{0}\rangle$ from the exact Bethe ansatz expression (\ref{psiR[y]}) with the rescaling above (\ref{lambda}). The third column is the result of Richardson extrapolation with exponent $\omega=1/2$ from the numerical values for system size $\leq L$, truncated at the error estimated by the extrapolation method. The exact asymptotics, given by (\ref{PhiF}) with four empty sets, is equal within $20$ digits to $0.69406512349927191436 - 
 0.36669029177597961516\,\rmi$.}
  \label{Table extrapolation}
\end{table}
\end{subsubsection}

\begin{subsubsection}{Step configurations}
In the case of unit step configurations $\mathcal{S}_{X/L}$, where sites from $X$ to $X+N-1$ are occupied while the rest of the system is empty, the density profile is called $\rho_{X/L}^{\mathcal{S}}$. The corresponding element of the eigenvector is a Vandermonde determinant. The calculation of its large $L$ asymptotics is significantly more involved \cite{P2015.2} than in the flat case, and can be obtained using two-dimensional Euler-Maclaurin formula in a triangular domain with combinations of logarithmic and square root singularities at all the edges and corners. From (\ref{R}), one has $\mathcal{R}[\rho_{x}^{\mathcal{S}}]=-\overline{\rho}(1-\overline{\rho})/2$, and the eigenvectors are  (see \ref{appendix Bethe})
\begin{equation}
\Phi_{r}[\rho_{x}^{\mathcal{S}}]=\rme^{-2\rmi\pi p_{r}(\overline{\rho}+x)}\Phi_{r}^{\mathcal{S}}\;
\end{equation}
and $\Phi_{r}[\tilde{\rho}_{x}^{\mathcal{S}}]=\rme^{2\rmi\pi p_{r}x}\Phi_{r}^{\mathcal{S}}$ for the reversed profile $\tilde{\rho}_{x}^{\mathcal{S}}$ according to (\ref{phiL[phiR]}). The quantity $\Phi_{r}^{\mathcal{S}}$ is equal to
\begin{eqnarray}
\label{PhiS}
&&\fl\hspace{15mm} \Phi_{r}^{\mathcal{S}}
=\frac{(\rmi\pi/2)^{m_{r}^{2}}}{(2\pi)^{m_{r}}}\,\omega(A_{0}^{+})\omega(A_{0}^{-})\omega(A^{+})\omega(A^{-})\omega(A_{0}^{+},A_{0}^{-})\omega(A^{+},A^{-})\\
&&\fl\hspace{42mm} \times\exp\Big(\lim_{\Lambda\to\infty}-m_{r}^{2}\log\Lambda+\int_{-\Lambda}^{2\pi c_{r}}\rmd u\,\frac{(\chi_{r}''(u))^{2}}{2}\Big)\;,\nonumber
\end{eqnarray}
with combinatorial factors
\begin{equation}
\label{omega(A)}
\omega(A)=\prod_{\scriptstyle a,a'\in A\atop\scriptstyle a>a'}(a-a')
\quad\text{and}\quad
\omega(A,A')=\prod_{a\in A}\prod_{a'\in A'}(a+a')\;.
\end{equation}
More generally, Richardson extrapolation of finite size Bethe ansatz numerics indicate that for any configuration $\mathcal{S}_{x}^{\rho_{-},\rho_{+}}$ corresponding at large scale to a step density profile $\rho_{0}$ with densities $\rho_{+}$ between $x$ and $x+a$ and $\rho_{-}$ elsewhere, $\rho_{-}<\rho_{+}$, one has
\begin{equation}
\label{PhiGeneralStep[kappa]}
\Phi_{r}[\rho_{0}]=\rme^{-2\rmi\pi p_{r}\kappa[\rho_{0}]}\Phi_{r}^{\mathcal{S}}\;,
\end{equation}
where $\kappa[\rho_{0}]=x+a$ is equal to the position of the center of the ramp in the moving frame at large times, defined more generally by (\ref{kappa}). This was checked for configurations of the form $(1^{r_{+}}0^{r-r_{+}})^{\ell_{+}L}(1^{r_{-}}0^{r-r_{-}})^{\ell_{-}L}$, for all the cases with $0\leq r_{-}<r_{+}\leq r\leq5$ and $\ell_{+}=\ell_{-}$, and for some other cases with $\ell_{+}=2\ell_{-}$ and $\ell_{+}=\ell_{-}/2$, $r_{-}<r_{+}$. An exact match was found with (\ref{PhiGeneralStep[kappa]}) within the error estimator of the extrapolation method.
\end{subsubsection}

\begin{subsubsection}{Generic configurations}
From section \ref{section Q deterministic}, any generic smooth density profile leads asymptotically to the same linear decreasing profile (\ref{rho(x,tau) large tau}) for large $t\sqrt{L}$, and the current fluctuations are expected to be the same as in the step case. Checking this with high precision using Richardson extrapolation does not seem possible, however, due to the lack of a natural sequence of configurations $\mathcal{C}_{L}$ leading to a clean expansion in powers of $1/\sqrt{L}$ for the eigenvectors. Nevertheless, limited numerics on linear and sinusoidal profiles seem to confirm the asymptotics (\ref{phi asymptotics[R,Phi]}) with $\mathcal{R}[\rho_{0}]$ given by (\ref{R}). These numerics also seem to indicate the presence of extra non-universal constants shifting the current, that have to be removed in order to recover (\ref{PhiGeneralStep[kappa]}). The non-universal constants seem to vanish for the left eigenvector with linear increasing profile and the right eigenvector with linear decreasing profile, which is probably related to the fact that the expression (\ref{rho(x,tau) large tau}) for the density profile at large time is exact for a linear increasing initial density profile, as in the case of a step profile, see \ref{appendix Burgers}.
\end{subsubsection}
\end{subsection}

\begin{subsection}{Probability distribution of the current fluctuations}
The probability density $P_{t}$ of the current fluctuations $\xi_{t}$ is obtained from Fourier transform of the generating function (\ref{G[c,Phi]}). One has
\begin{equation}
\label{P[G]}
P_{t}(u)=\int_{-\infty}^{\infty}\frac{\rmd s}{2\pi}\,\rme^{-\rmi su}G_{t}(\rmi s)\;.
\end{equation}
The probability density is plotted in the flat $\to$ flat case in figure \ref{Fig Pt}, based on numerical evaluations where only a finite number of eigenstates $r$ are kept in (\ref{G[c,Phi]}). More eigenstates are needed to ensure reasonable convergence for small values of $t$.
\begin{figure}
  \begin{center}
      \begin{tabular}{c}
        \includegraphics[width=100mm]{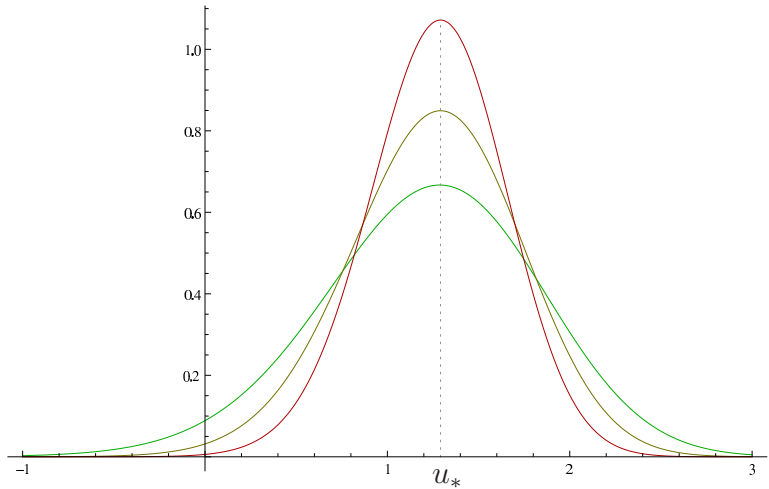}
        \begin{picture}(0,0)
          \put(-45.5,-1){$u_{*}$}
        \end{picture}
      \end{tabular}
  \end{center}
  \caption{Plots as a function of $u$ of the probability density $t^{1/3}P_{t}(t^{1/3}u)$ of rescaled current fluctuations $\xi_{t}/t^{1/3}$, computed numerically from (\ref{P[G]}), (\ref{G flat flat}) in the flat $\to$ flat case. The integral over $s$ in (\ref{P[G]}) is discretized as a sum over $101$ equally spaced values between $-5/t$ and $5/t$. The generating function $G$ is evaluated from (\ref{G flat flat}) by summing over the $12$ eigenstates corresponding to sets $A\equiv A_{0}^{+}=A^{-}$ and $\overline{A}=A_{0}^{-}=A^{+}$ with $\sum_{a\in A}a+\sum_{a\in\overline{A}}a\leq4$. The three graphs correspond respectively to rescaled time $t=0.2$ (flattest curve), $t=0.1$ and $t=0.05$ (most peaked curve). When $t\to0$, the random variable $\xi_{t}/t^{1/3}$ converges with probability $1$ to $u_{*}\approx1.291468$, indicated by a dotted line.}
  \label{Fig Pt}
\end{figure}

Making the change of variables $s\to c$ removes the necessity to solve (\ref{c(s)}) in the expression (\ref{P[G]}), (\ref{G flat flat}) for the probability density. One finds
\begin{equation}
\label{P[Phi,Phi]}
P_{t}(u)=\frac{1}{\mathcal{Z}_{t}}\int_{\rme^{-\rmi\pi/3}\infty}^{\rme^{\rmi\pi/3}\infty}\frac{\rmd c}{\rmi\sqrt{2\pi}}\,\rme^{2\pi c}\sum_{r}\rme^{-u\chi_{r}'(2\pi c)}\rme^{t\chi_{r}(2\pi c)}\,\Phi_{r}[\tilde{\rho}_{\text{f}}]\,\Phi_{r}[\rho_{\text{i}}]\;,
\end{equation}
with $c_{r}=c_{r}(\rmi s)$ replaced by $c$ in the expressions for $\Phi_{r}$. The integration range follows from the large $|s|$ asymptotics $2\pi c_{r}(s)\simeq\big(\frac{3\pi s}{2\sqrt{2}}\big)^{2/3}$ for $s\not\in\mathbb{R}^{-}$. The quantities $c_{r}(\rmi s)$, $s\in\mathbb{R}$ are plotted in figure \ref{Fig c} for some eigenstates.
\begin{figure}
  \begin{center}
    \begin{tabular}{c}\includegraphics[width=100mm]{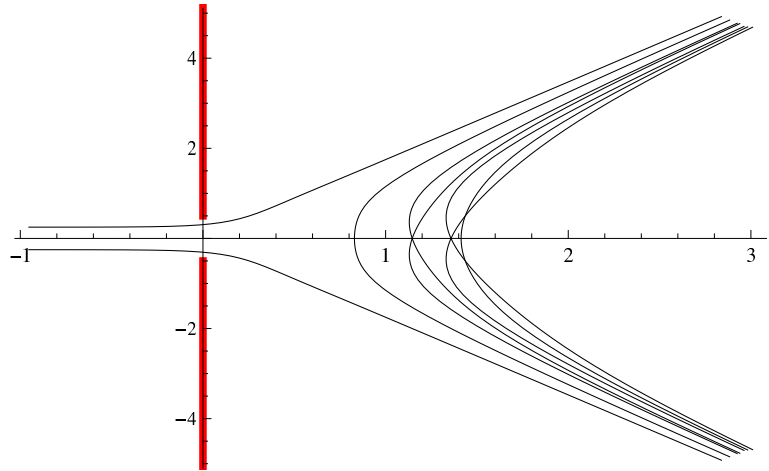}\end{tabular}
  \end{center}
  \caption{Plots in the complex plane of the curves $c_{r}(\rmi s)$, $s\in\mathbb{R}$ for the few first eigenstates $r$ contributing to the current fluctuations in the flat $\to$ flat case. The curve for the stationary state $r=0$ goes to $-\infty$ when $s\to0$. The curves for all the other eigenstates verify $\Re\,c_{r}(\rmi s)>0$. When $s\to\pm\infty$, the curves diverge to $\rme^{\pm\rmi\pi/3}\infty$. The thick, red lines correspond to the branch cuts $\rmi[\half,\infty)$ and $-\rmi[\half,\infty)$ in the variable $c$ of the integrand in (\ref{P flat flat}).}
  \label{Fig c}
\end{figure}
\end{subsection}
\end{section}

\begin{section}{Large deviations in the flat $\to$ flat case}
\label{section flat -> flat}
In this section, we specialize to the case of an evolution conditioned on flat initial and final configurations. There, the sum over eigenstates in the probability density of current fluctuations can be computed explicitly as an infinite product. From this expression, an exact formula is derived for the large deviations of current fluctuations at short rescaled time $t$.

\begin{subsection}{Probability distribution}
If both $\rho_{i}$ and $\rho_{f}$ correspond to flat profiles of density $\overline{\rho}$, the generating function (\ref{G[c,Phi]}) simplifies to
\begin{equation}
\label{G flat flat}
G_{t}(s)=\frac{1}{\mathcal{Z}_{t}}\sum_{r}
\openone_{\{A_{0}^{+}=A^{-}\}}
\openone_{\{A_{0}^{-}=A^{+}\}}
\frac{\rme^{2\pi c_{r}}\rme^{t\chi_{r}(2\pi c_{r})}}{\sqrt{2\pi}\sqrt{1+\rme^{2\pi c_{r}}}\,\chi_{r}''(2\pi c_{r})}\;.
\end{equation}
The probability density of current fluctuations (\ref{P[Phi,Phi]}) is then equal to
\begin{eqnarray}
&& P_{t}(u)=\frac{1}{\mathcal{Z}_{t}}\int_{\rme^{-\rmi\pi/3}\infty}^{\rme^{\rmi\pi/3}\infty}\frac{\rmd c}{\rmi\sqrt{2\pi}}\,\frac{\rme^{2\pi c}\rme^{-u\chi_{0}'(2\pi c)}\rme^{t\chi_{0}(2\pi c)}}{\sqrt{1+\rme^{2\pi c}}}\sum_{A_{0}^{\pm}\in\mathbb{N}+\half}\openone_{\{|A_{0}^{+}|=|A_{0}^{-}|\}}\nonumber\\
&&\hspace{35mm} \times\rme^{4\sqrt{\pi}u\big(\sum_{a\in A_{0}^{+}}\sqrt{-\rmi}\sqrt{a-\rmi c}+\sum_{a\in A_{0}^{-}}\sqrt{\rmi}\sqrt{a+\rmi c}\big)}\\
&&\hspace{35mm} \times\rme^{-\frac{16\pi^{3/2}t}{3}\big(\sum_{a\in A_{0}^{+}}\sqrt{\rmi}\,(a-\rmi c)^{3/2}+\sum_{a\in A_{0}^{-}}\sqrt{-\rmi}\,(a+\rmi c)^{3/2}\big)}\;,\nonumber
\end{eqnarray}
where $\chi_{0}$ is the function (\ref{chi[A,zeta]}) for the stationary state, corresponding to four empty sets. The sum over the sets can be computed by adding a contour integral to enforce the constraint $|A_{0}^{+}|=|A_{0}^{-}|$. One has
\begin{eqnarray}
\label{P flat flat}
&& P_{t}(u)=\frac{1}{\mathcal{Z}_{t}}\int_{\rme^{-\rmi\pi/3}\infty}^{\rme^{\rmi\pi/3}\infty}\frac{\rmd c}{\rmi\sqrt{2\pi}}\,\frac{\rme^{2\pi c}\rme^{-u\chi_{0}'(2\pi c)}\rme^{t\chi_{0}(2\pi c)}}{\sqrt{1+\rme^{2\pi c}}}\oint\frac{\rmd z}{2\rmi\pi z}\nonumber\\
&&\hspace{25mm} \prod_{a\in\mathbb{N}+\half}\Big[\Big(1+z\,\rme^{4\sqrt{-\rmi}\sqrt{\pi}u\sqrt{a-\rmi c}-\frac{16\sqrt{\rmi}\,\pi^{3/2}t}{3}(a-\rmi c)^{3/2}}\Big)\\
&&\hspace{40mm} \Big(1+z^{-1}\,\rme^{4\sqrt{\rmi}\sqrt{\pi}u\sqrt{a+\rmi c}-\frac{16\sqrt{-\rmi}\,\pi^{3/2}t}{3}(a+\rmi c)^{3/2}}\Big)\Big]\;.\nonumber
\end{eqnarray}
\end{subsection}

\begin{subsection}{First cumulants of the current}
The first cumulants of $\xi_{t}$ are obtained by taking derivatives with respect to $s$ at $s=0$ of the generating function $G_{t}(s)$ (\ref{G flat flat}). They can be computed numerically by truncating the sum over first eigenstates to keep only the eigenstates corresponding to small values of $\sum_{a\in A_{0}^{+}}a+\sum_{a\in A_{0}^{-}}a+\sum_{a\in A^{+}}a+\sum_{a\in A^{-}}a$ as the sum over eigenstates converges rather quickly if the rescaled time $t$ is not too small. This is especially true when either the initial or the final state is flat: then most eigenstates do not contribute because of the constraints $A_{0}^{+}=A^{-}$ and $A_{0}^{-}=A^{+}$ on the sets. The four first cumulants are plotted in the flat $\to$ flat case as a function of time in figure \ref{Fig cumulants flat flat}. The Derrida-Appert ratio $\langle\xi_{t}^{2}\rangle_{\text{c}}\langle\xi_{t}^{4}\rangle_{\text{c}}/\langle\xi_{t}^{3}\rangle_{\text{c}}^{2}$ from \cite{DA1999.1} is also plotted in figure \ref{Fig DA ratio}. It has a non-zero finite limit both when $t\to0$ and $t\to\infty$, unlike the individual cumulants.
\begin{figure}
  \begin{center}
    \begin{tabular}{c}\includegraphics[width=150mm]{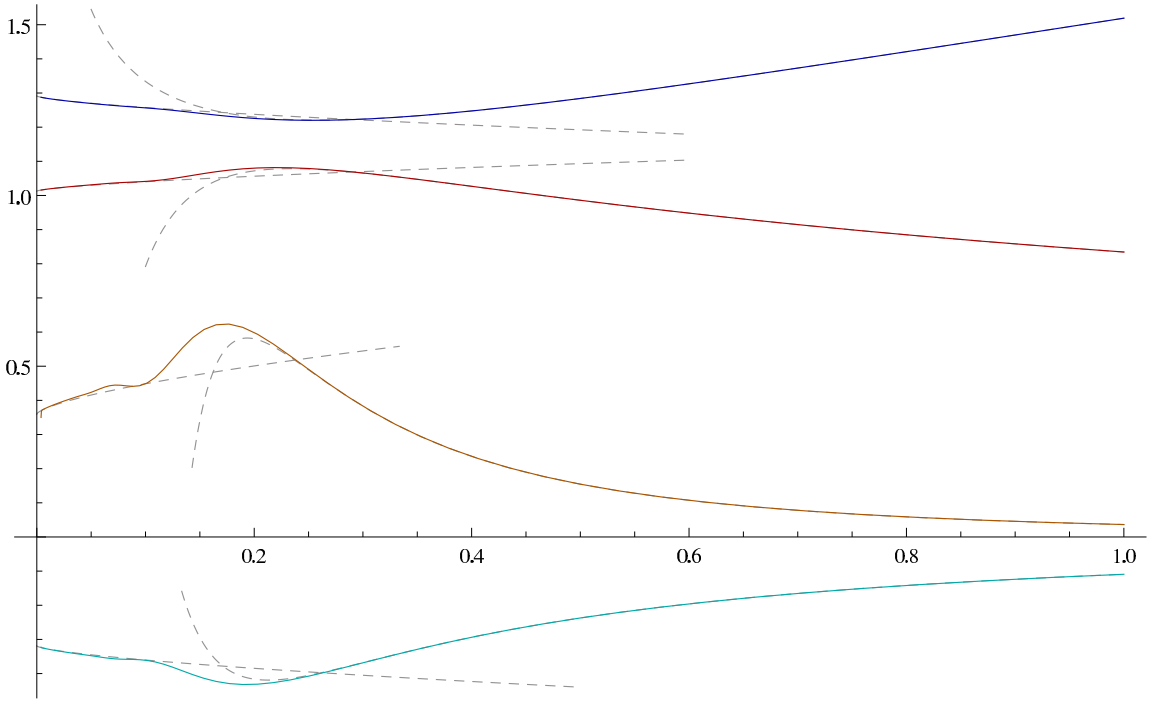}\end{tabular}
  \end{center}
  \caption{Rescaled first cumulants $\langle\xi_{t}^{k}\rangle_{\text{c}}/t^{k-2/3}$ of the current fluctuations $\xi_{t}$ plotted as a function of $t$, for an evolution conditioned on flat initial and final states. The solid lines correspond to values of the cumulants obtained by taking derivatives at $s=0$ of the generating function (\ref{G flat flat}), computed by summing over the $6639349$ eigenstates corresponding to sets $A\equiv A_{0}^{+}=A^{-}$ and $\overline{A}=A_{0}^{-}=A^{+}$ with $\sum_{a\in A}a+\sum_{a\in\overline{A}}a\leq60$. From top to bottom, they represent the rescaled average (blue), variance (red), fourth cumulant (orange) and third cumulant (cyan). The dashed lines represent the short and long time values, given respectively by (\ref{cumulants small t}) and (\ref{cumulants large t}).}
  \label{Fig cumulants flat flat}
\end{figure}
\begin{figure}
  \begin{center}
    \begin{tabular}{c}\includegraphics[width=100mm]{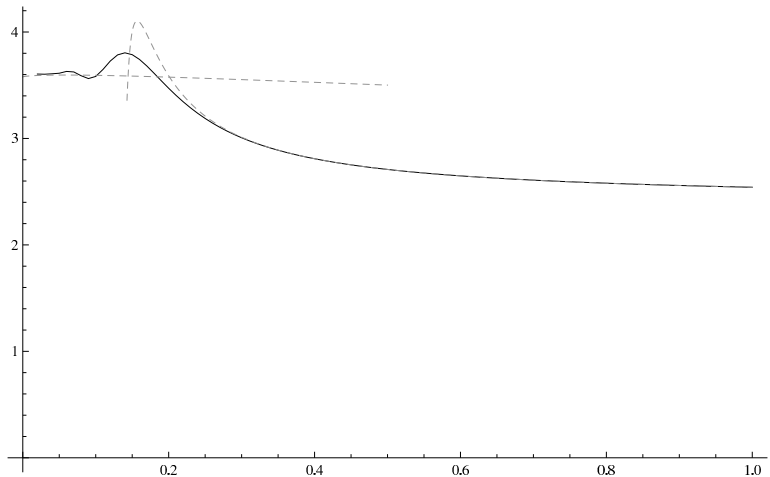}\end{tabular}
  \end{center}
  \caption{Derrida-Appert ratio $\langle\xi_{t}^{2}\rangle_{\text{c}}\langle\xi_{t}^{4}\rangle_{\text{c}}/\langle\xi_{t}^{3}\rangle_{\text{c}}^{2}$ plotted as a function of the rescaled time $t$ for an evolution conditioned on flat initial and final states, along with large and small $t$ asymptotics from (\ref{cumulants large t}) and (\ref{cumulants small t}).}
  \label{Fig DA ratio}
\end{figure}
\end{subsection}

\begin{subsection}{Large deviations of the current in the long time limit}
At large time $t$, the generating function (\ref{G flat flat}) is essentially equal to the contribution of the stationary state corresponding to four empty sets, which implies
\begin{equation}
\label{LDF f st}
\langle\rme^{s\xi_{t}}\rangle\simeq\rme^{tf_{\text{st}}(s)}\;,
\end{equation}
with
\begin{equation}
\label{fst[chi0]}
f_{\text{st}}(s)=\chi_{0}(2\pi c_{0}(s))\;
\end{equation}
and $c_{0}(s)$ solution of $\chi_{0}'(2\pi c_{0}(s))=s$. The function $f_{\text{st}}$ is the stationary state cumulant generating function characteristic of KPZ universality at large time. It corresponds to cumulants of the current scaling as $\langle\xi_{t}^{k}\rangle_{\text{c}}\sim t$ in the long time limit. It was first obtained for periodic TASEP in \cite{DL1998.1} by Derrida and Lebowitz, see also \cite{DA1999.1}, and was subsequently derived for other models in the KPZ universality class: ASEP \cite{LK1999.1}, open TASEP on the transition line between low/high density phase and maximal current phase \cite{LM2011.1,GLMV2012.1,L2015.1}, discrete time ASEP with parallel update \cite{PM2006.1}, a directed polymer model \cite{BD2000.1}, and the asymmetric avalanche process \cite{PPH2003.1}. The stationary large deviation function was extended to the crossover between KPZ and equilibrium fluctuations in ASEP with weak asymmetry \cite{ADLvW2008.1,PM2009.1,P2010.1,S2011.1}. Some results were also obtained for the average of current fluctuations in ASEP with several species of particles \cite{DE1999.1,C2008.1}.

At large time $T$, the total current $Q$ for periodic TASEP is equal with probability $1$ to $N(L-N)T/(L-1)$ at leading order in $T$, see \textit{e.g.} \cite{D1998.1}. It implies $\xi_{t}\to1$ when $t\to\infty$ for the current fluctuations. The Legendre transform $g_{\text{st}}(u)=\max_{s}(us-f_{\text{st}}(s))$ of $f_{\text{st}}$ describes the probability of rare events when $\xi_{t}\simeq tu$ with $u\neq1$ at large $t$. It is known as the \textit{large deviation function} of the current for the stationary state, and verifies
\begin{equation}
\label{LDF g st}
P(\xi_{t}=tu)\simeq\rme^{-tg_{\text{st}}(u)}\;.
\end{equation}
In the notations of \cite{DA1999.1}, one has $f_{\text{st}}(s)=G(s\sqrt{2\pi})/\sqrt{2\pi}$, $g_{\text{st}}(u)=-H(u)/\sqrt{\pi}$, and $g_{\text{st}}$ behaves for large argument as $g_{\text{st}}(u)\simeq2\sqrt{3}u^{5/2}/(5\pi)$ when $u\to\infty$ and $g_{\text{st}}(u)\simeq4|u|^{3/2}/3$ when $u\to-\infty$.

The functions $f_{\text{st}}$ and $g_{\text{st}}$ do not contain any information about the initial and the final state of the evolution. Some information about the evolution can however be found in the first order correction in $t$. In the flat $\to$ flat case, one obtains from (\ref{G flat flat}) $\langle\rme^{s\xi_{t}}\rangle\simeq\rme^{tf_{\text{st}}^{\text{f}\to\text{f}}(s)}$ with
\begin{equation}
\fl\hspace{10mm} f_{\text{st}}^{\text{f}\to\text{f}}(s)\simeq f_{\text{st}}(s)
+\frac{2\pi c_{0}(s)-\log\sqrt{2\pi}-\log\sqrt{1+\rme^{2\pi c_{0}(s)}}-\log\chi_{0}''(2\pi c_{0}(s))}{t}\;
\end{equation}
up to exponentially small corrections in $t$. For the first cumulants of the current, it leads to
\begin{eqnarray}
\label{cumulants large t}
&&\hspace{-5mm} \langle\xi_{t}\rangle\simeq t
+\Big(1-\frac{1}{\sqrt{2}}\Big)\sqrt{\pi}\nonumber\\
&&\hspace{-5mm} \langle\xi_{t}^{2}\rangle_{\text{c}}\simeq\frac{\sqrt{\pi}}{2}\,t
+\Big(3-\frac{1}{\sqrt{2}}-\frac{4}{\sqrt{3}}\Big)\pi\\
&&\hspace{-5mm} \langle\xi_{t}^{3}\rangle_{\text{c}}\simeq\Big(\frac{3}{2}-\frac{8}{3^{3/2}}\Big)\pi t
+\Big(11+\frac{2\sqrt{2}}{\sqrt{3}}+\frac{5}{\sqrt{2}}-\frac{28}{\sqrt{3}}\Big)\pi^{3/2}\nonumber\\
&&\hspace{-5mm} \langle\xi_{t}^{4}\rangle_{\text{c}}\simeq\Big(\frac{15}{2}+\frac{9}{\sqrt{2}}-\frac{24}{\sqrt{3}}\Big)\pi^{3/2}t
+\Big(\frac{319}{3}+\frac{20\sqrt{2}}{\sqrt{3}}+\frac{93}{\sqrt{2}}-84\sqrt{3}-\frac{96}{\sqrt{5}}\Big)\pi^{2}\;.\nonumber
\end{eqnarray}
These expressions are plotted in figure \ref{Fig cumulants flat flat} along with the exact finite time values of the cumulants.
\end{subsection}

\begin{subsection}{Large deviations of the current in the short time limit}
At short time, the first cumulants of $\xi_{t}$ scale as $\langle\xi_{t}^{k}\rangle_{\text{c}}\sim t^{k-2/3}$ in the flat $\to$ flat case, as seen in figure \ref{Fig cumulants flat flat}. It corresponds for the generating function to the behaviour
\begin{equation}
\label{LDF f short t}
\langle\rme^{s\xi_{t}/t}\rangle\sim\rme^{t^{-2/3}f_{0}^{\text{f}\to\text{f}}(s)}\;,
\end{equation}
which is related by Legendre transform $f_{0}^{\text{f}\to\text{f}}(s)=\max_{u}(su-g_{0}^{\text{f}\to\text{f}}(u))$ to the large deviations
\begin{equation}
\label{LDF g short t}
P(\xi_{t}=t^{1/3}u)\sim\rme^{-t^{-2/3}g_{0}^{\text{f}\to\text{f}}(u)}\;.
\end{equation}
This kind of short-time large deviations were already observed in \cite{LK2006.1} as a consequence of the fact that spatial correlations scale as $T^{2/3}$ for small $t$. They can be understood by breaking up the full system into around $L/T^{2/3}\sim t^{-2/3}$ almost stationary subsystems and by using stationary-like large deviations for each subsystem of size $T^{2/3}$ \cite{LK2006.1}.

The function $g_{0}^{\text{f}\to\text{f}}$ can be computed explicitly from a saddle point analysis of the exact formula (\ref{P flat flat}) for the probability at time $t$ of $\xi_{t}$, see \ref{appendix saddle point}. One finds
\begin{equation}
\label{P[Z] flat flat asymptotics t0}
t^{1/3}P_{t}(t^{1/3}u)\simeq\frac{t^{1/3}\rme^{t^{-2/3}\Xi(u)}}{\sqrt{2\pi}\mathcal{Z}_{t}}\;
\end{equation}
with
\begin{equation}
\label{Xi[int,Ai]}
\boxed{\boxed{\;
\Xi(u)
=\int_{\rme^{-\rmi\theta}\infty}^{\rme^{\rmi\theta}\infty}\frac{\rmd w\,w}{2^{5/3}\rmi\pi}\log\big(1-\rme^{\frac{w^{3}}{3}+2^{2/3}uw}\big)
=\sum_{k=1}^{\infty}\frac{\Ai'(-2^{2/3}k^{2/3}u)}{2^{2/3}k^{5/3}}\;
}}
\end{equation}
in the range $u\in[0,(3\pi/2)^{2/3}]$. Outside of this interval, one has to add $3u/2$ to the expressions for $\Xi$ in (\ref{Xi[int,Ai]}) when $u<0$, and to add $\text{a.c.(u)}$ defined in (\ref{a.c.+(u)}) for $u>(3\pi/2)^{2/3}$. These additional terms make the function $\Xi$ analytic around the whole real axis. The function $\Xi$ is plotted in figure \ref{Fig Xi}. The path of integration in (\ref{Xi[int,Ai]}), required to avoid the branch cuts due to the logarithm plotted in figure \ref{Fig cuts w}, goes to infinity in directions specified by angles $\pm\theta$, $\pi/6<\theta<\pi/2$. The symbol $\Ai$ denotes the Airy function.
\begin{figure}
  \begin{center}
    \begin{tabular}{c}
      \includegraphics[width=150mm]{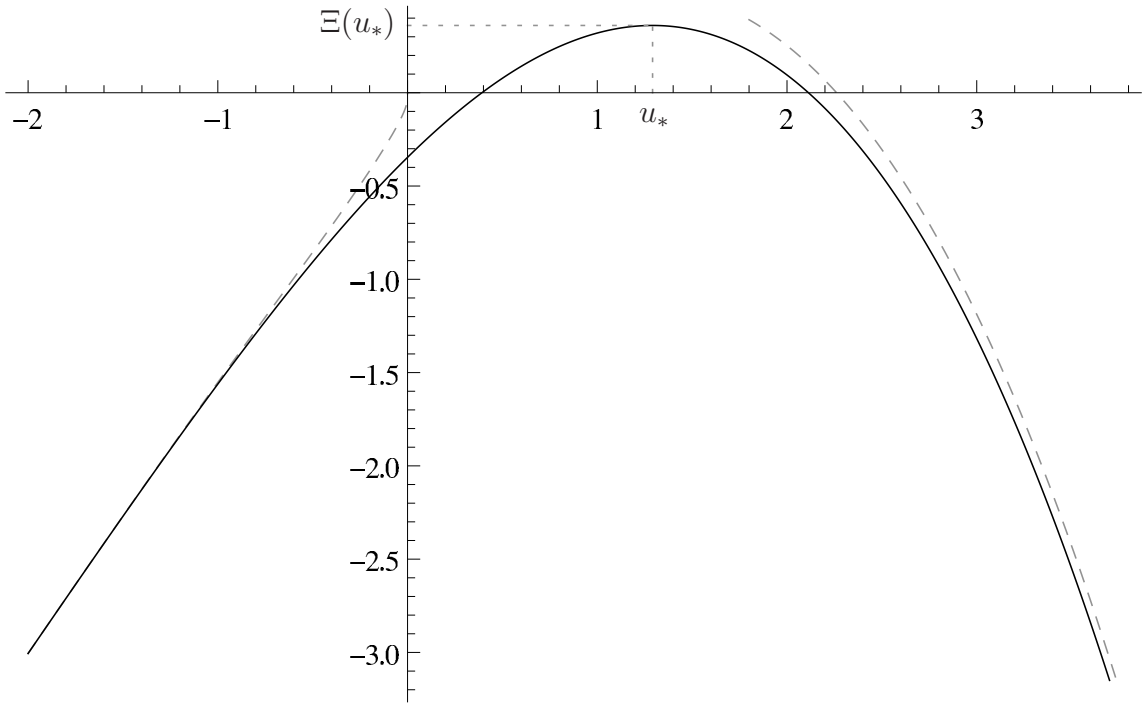}
      \begin{picture}(0,0)
        \put(-68,77){$u_{*}$}
        \put(-110,89){$\Xi(u_{*})$}
      \end{picture}
    \end{tabular}
  \end{center}
  \caption{Graph of the function $\Xi$ (\ref{Xi[int,Ai]}), related to the large deviation function of the current at short time (\ref{LDF g short t}), (\ref{g0[Xi]}) for an evolution conditioned on flat initial and final states. The asymptotics (\ref{Xi(-infty)}) and (\ref{Xi(+infty)}) are plotted with dashed lines. The dotted lines indicate the maximum of $\Xi$.}
  \label{Fig Xi}
\end{figure}
\begin{figure}
  \begin{center}
    \begin{tabular}{cc}
      \includegraphics[width=75mm]{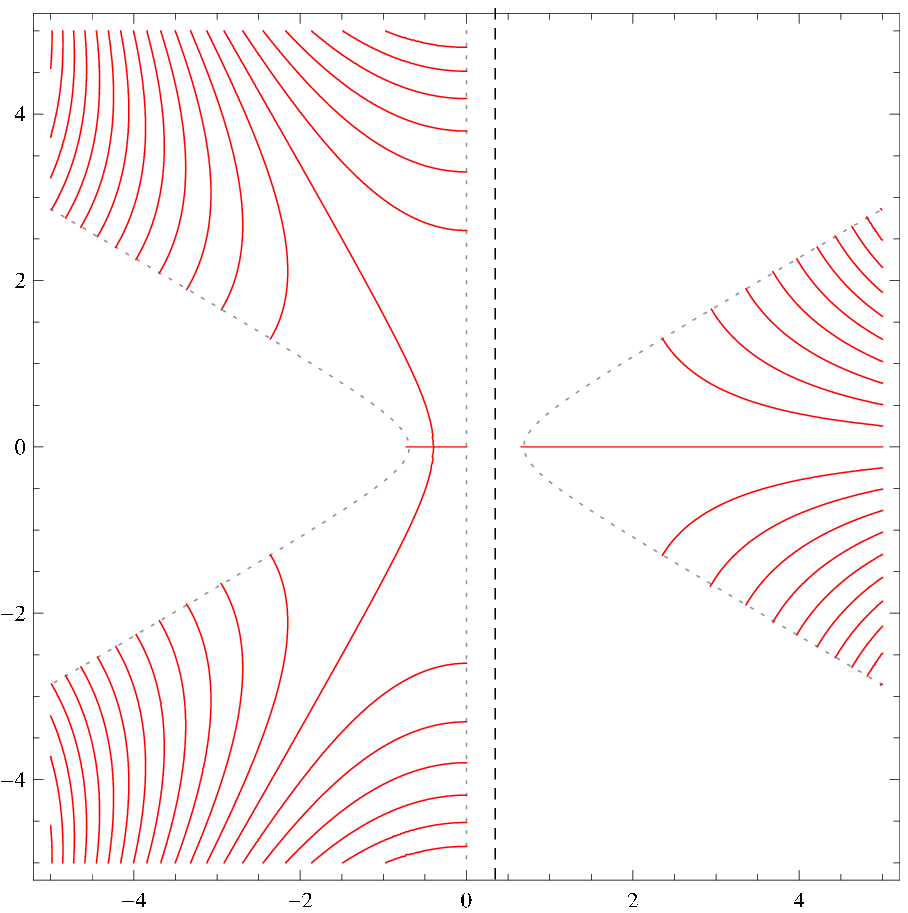}
      &
      \includegraphics[width=75mm]{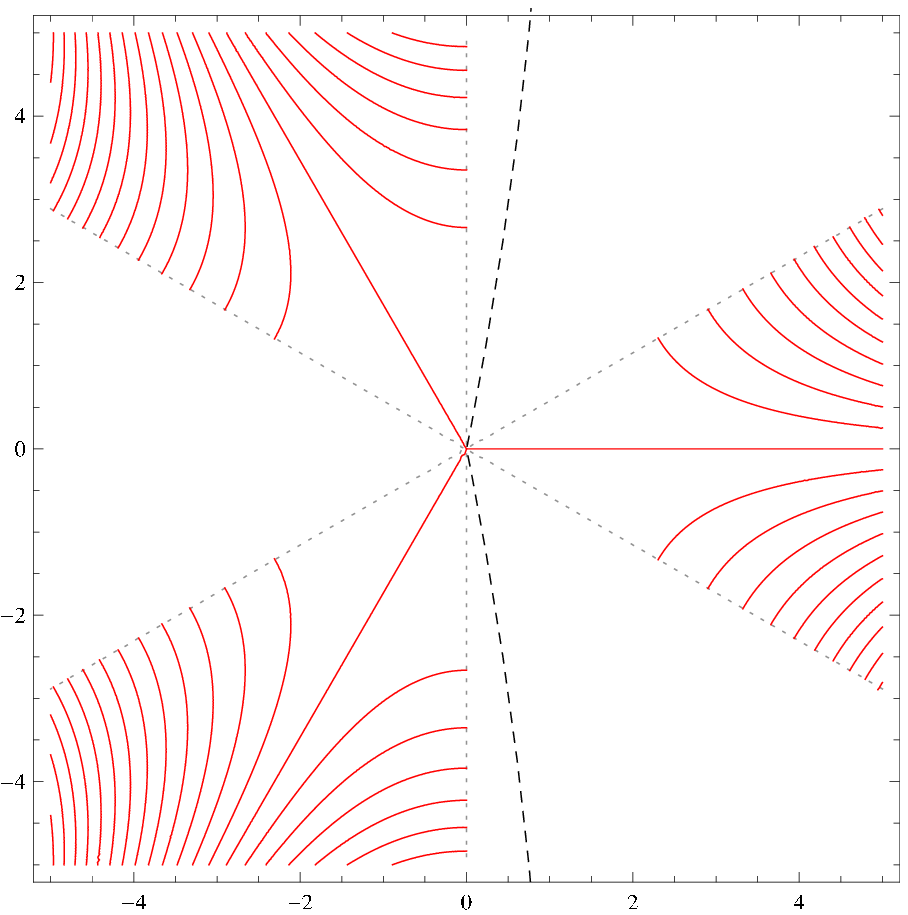}\\
      \includegraphics[width=75mm]{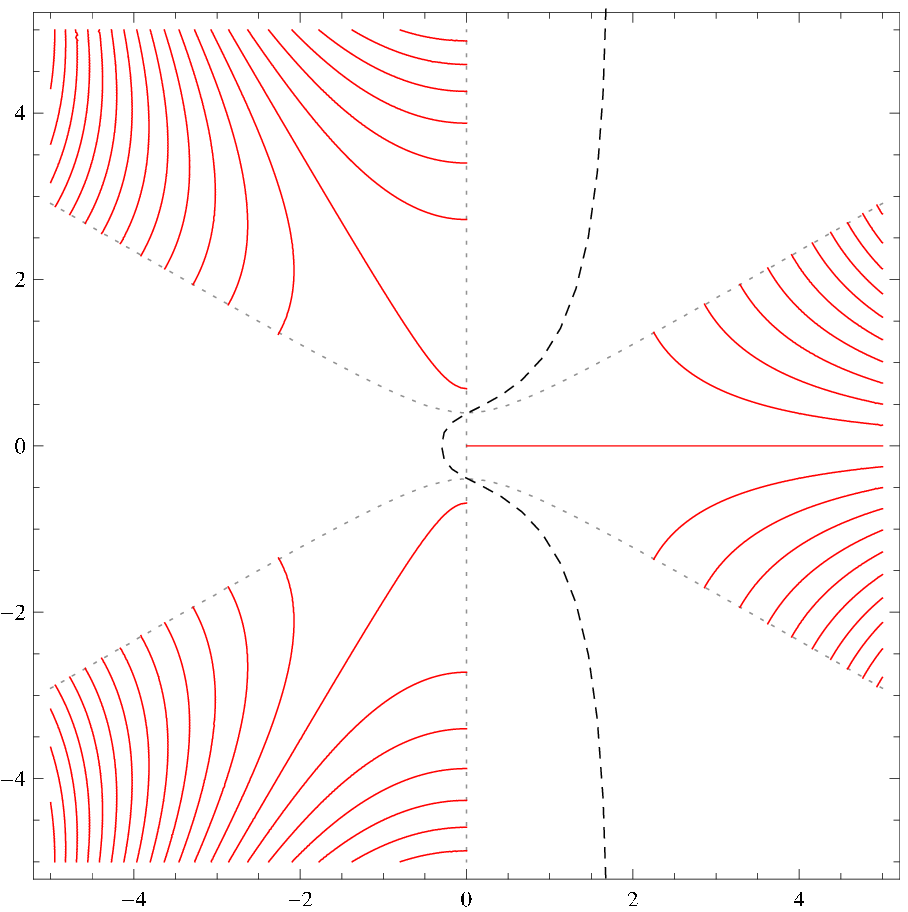}
      &
      \includegraphics[width=75mm]{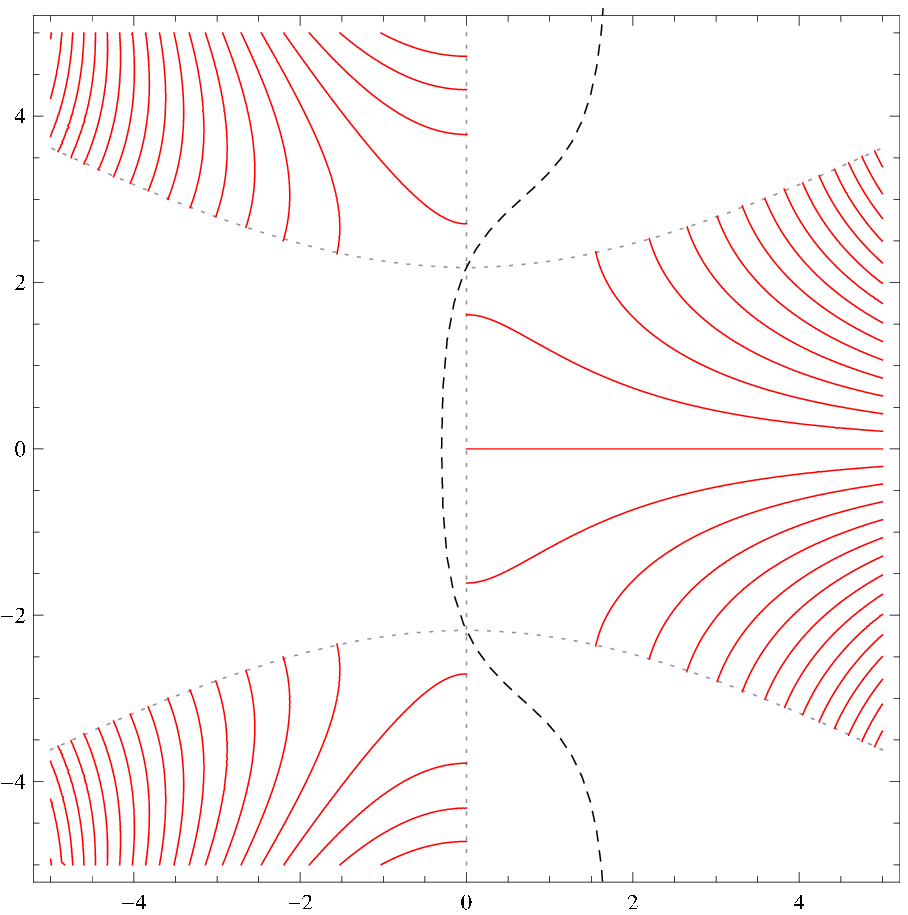}
    \end{tabular}
  \end{center}
  \caption{Branch cuts of the integrand in the expression (\ref{Xi[int,Ai]}) for the quantity $\Xi(u)$ for various values of the parameter $u$. The solid red curves represent the branch cuts of $\log\big(1-\rme^{\frac{w^{3}}{3}+2^{2/3}uw}\big)$ as a function of $w$. The dotted grey curves are the locus of the $w$ such that $\Re(\frac{w^{3}}{3}+2^{2/3}uw)=0$. The dashed black curves correspond to a suitable contour of integration oriented upward for (\ref{Xi[int,Ai]}). The different graphs correspond from left to right to $u$ equal to $-0.1$, $0$ (top row), $0.1$, $3$ (bottom row).}
  \label{Fig cuts w}
\end{figure}

Calling $u_{*}\approx1.29146805131163785850008244580$ the location of the maximum of $\Xi$, the normalization condition $\int_{-\infty}^{\infty}\rmd u\,t^{1/3}P_{t}(t^{1/3}u)=1$ implies the small $t$ asymptotics for the ratio between the probability to observe the system in the flat configuration $\mathcal{F}$ at time $t$ starting from $\mathcal{F}$ and the stationary probability of $\mathcal{F}$. One has
\begin{equation}
\label{Z asymptotics}
\mathcal{Z}_{t}\simeq\frac{t^{2/3}\rme^{t^{-2/3}\Xi(u_{*})}}{\sqrt{-\Xi''(u_{*})}}\;.
\end{equation}
This was checked numerically by truncating the sum over all eigenstates, see figure \ref{Fig Z}.
\begin{figure}
  \begin{center}
    \begin{tabular}{c}\includegraphics[width=100mm]{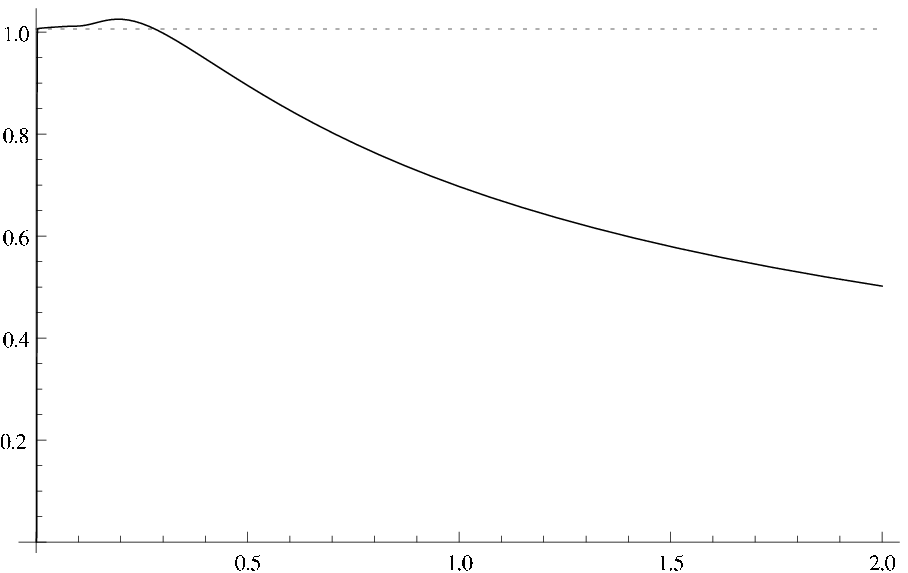}\end{tabular}
  \end{center}
  \caption{Normalization factor $\mathcal{Z}_{t}$, equal to the probability to observe the system in a flat configuration $\mathcal{F}$ for an evolution starting in $\mathcal{F}$, divided by the stationary probability $1/\Omega$ of $\mathcal{F}$. The normalization is evaluated by summing over the $6639349$ eigenstates corresponding to sets $A\equiv A_{0}^{+}=A^{-}$ and $\overline{A}=A_{0}^{-}=A^{+}$ with $\sum_{a\in A}a+\sum_{a\in\overline{A}}a\leq60$. The quantity $t^{-2/3}\rme^{-t^{-2/3}\Xi(u_{*})}\mathcal{Z}_{t}$ is plotted as a function of the rescaled time $t$ (solid line). The dotted line is the limit $t\to0$, equal to $(-\Xi''(u_{*}))^{-1/2}\approx1.006299331334999730981973520718$.}
  \label{Fig Z}
\end{figure}

The short time large deviation function from (\ref{LDF g short t}) is then equal to
\begin{equation}
\label{g0[Xi]}
g_{0}^{\text{f}\to\text{f}}(u)=\Xi(u_{*})-\Xi(u)\;,
\end{equation}
with $\Xi(u_{*})=\max_{u\in\mathbb{R}}\Xi(u)\approx0.360699035681939348898709742128$. The functions $g_{0}^{\text{f}\to\text{f}}$ (\ref{g0[Xi]}) and $g_{\text{st}}$, Legendre transform of (\ref{fst[chi0]}) (see \cite{DA1999.1} for technical details about the required analytic continuation), are plotted in figure \ref{Fig LDF}.
\begin{figure}
  \begin{center}
    \begin{tabular}{c}\includegraphics[width=100mm]{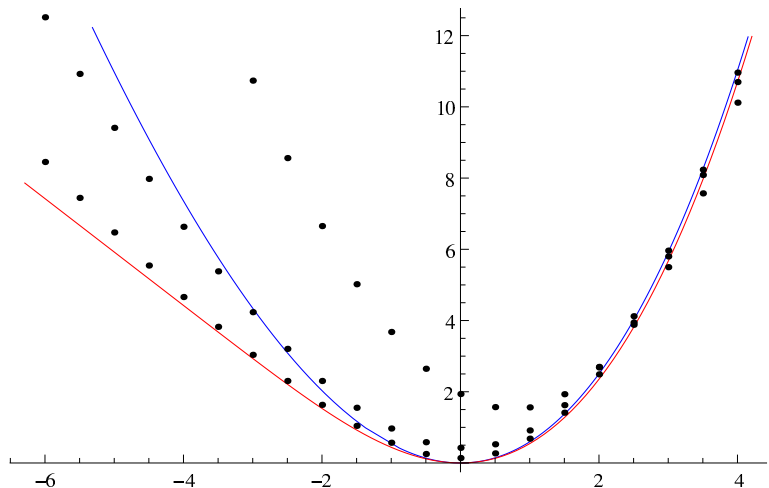}\end{tabular}
  \end{center}
  \caption{Graphs of shifted large deviation functions of the current as a function of their argument $u$. The solid red, lower curve represents the large deviation function at short time $g_{0}^{\text{f}\to\text{f}}(u_{*}+u)$ for an evolution conditioned on flat initial and final states, defined in (\ref{g0[Xi]}). The solid blue, upper curve represents the stationary large deviation function $g_{\text{st}}(1+u)$ plotted from its Legendre transform (\ref{fst[chi0]}). The shifts of $u_{*}\approx1.291468$ and $1$ are such that the minimum of both curves is located at $0$. The black dots correspond to finite time evaluations at $t=2$ (upper dots), $t=0.5$ and $t=0.2$ (lower dots) of $-t^{2/3}\log(t^{1/3}P_{\xi}(t^{1/3}(u+u_{*})))$ from (\ref{P[G]}), after discretizing the integral over $s$ and truncating the infinite sum over eigenstates in (\ref{G flat flat}).}
  \label{Fig LDF}
\end{figure}

The location $u_{*}$ of the maximum of $\Xi$ is the deterministic limit of the random variable $t^{-1/3}\xi_{t}$ in the limit $t\to0$, and $\langle\xi_{t}\rangle\simeq u_{*}\,t^{1/3}$. Higher cumulants can be computed numerically from (\ref{g0[Xi]}) as well, and the leading correction obtained from (\ref{P[Z] flat flat asymptotics t0}), (\ref{Z asymptotics}). One finds
\begin{eqnarray}
\label{cumulants small t}
&&\fl\hspace{10mm} \langle\xi_{t}/t\rangle\simeq 1.2914680513116378585\,t^{-2/3} - 0.15751847779536355747\nonumber\\
&&\fl\hspace{10mm} \langle(\xi_{t}/t)^{2}\rangle_{\text{c}}\simeq 1.0126383442452675715\,t^{-2/3} + 0.12793113905830599461\\
&&\fl\hspace{10mm} \langle(\xi_{t}/t)^{3}\rangle_{\text{c}}\simeq -0.31901850108546379645\,t^{-2/3} - 0.19190248943176059601\nonumber\\
&&\fl\hspace{10mm} \langle(\xi_{t}/t)^{4}\rangle_{\text{c}}\simeq 0.35959857102590971120\,t^{-2/3} + 0.41299282842296155443\;.\nonumber
\end{eqnarray}
These asymptotics are plotted in figure \ref{Fig cumulants flat flat} along with exact numerical values of the cumulants.

The asymptotics of the function $\Xi$ when its argument becomes large can be calculated explicitly. At large $|u|$, the expression (\ref{Xi[int,Ai]}) of $\Xi(u)$ is negligible compared to the extra terms $3u/2$ and $\text{a.c.(u)}$ required for the analytic continuation. When $u\to-\infty$, the asymptotics of the Airy function gives
\begin{equation}
\label{Xi(-infty)}
\Xi(u)\underset{u\to-\infty}{\simeq}\frac{3u}{2}-\frac{(-u)^{1/4}\rme^{-4(-u)^{3/2}/3}}{2\sqrt{2\pi}}\;.
\end{equation}
When $u\to\infty$, the sum $\text{a.c.(u)}$ (\ref{a.c.+(u)}) becomes an integral, that can be computed explicitly. Adding the first Euler-Maclaurin correction, one finds
\begin{equation}
\label{Xi(+infty)}
\Xi(u)\underset{u\to\infty}{\simeq}-\frac{2\sqrt{3}u^{5/2}}{5\pi}+\frac{3u}{4}+\mathcal{O}(u^{-1/2})\;.
\end{equation}
We observe that the short and long time large deviations $g_{0}^{\text{f}\to\text{f}}(u)$ and $g_{\text{st}}(u)$ have the same asymptotics $\sim u^{5/2}$ with the same coefficient in front when $u\to\infty$, see also figure \ref{Fig LDF}. Similarities between short and long time large deviations were already observed from simulations in \cite{LK2006.1}. For $u\to-\infty$ on the other hand, $g_{0}^{\text{f}\to\text{f}}(u)$ grows as $|u|$, slower than $g_{\text{st}}(u)$ which grows as $|u|^{3/2}$.
\end{subsection}

\begin{subsection}{Comparison with an evolution not conditioned on the final state: simulations}
Exact results for an evolution conditioned only on the initial state are still out of reach since they would require the large $L$ asymptotics of the sum over all configurations $\sum_{\mathcal{C}}\langle\mathcal{C}|\phi_{r}\rangle$ for the first eigenstates, which is not known yet. It is however possible to study current fluctuations from simulations when the evolution is not conditioned on the final state. The first cumulants of the current are studied from simulations of periodic TASEP with $N=1000$ particles on $L=2000$ sites and flat initial state. They are plotted along with the flat $\to$ flat exact results in figure \ref{Fig cumulants flat stat}. We use in this section the superscript $\text{f}\to$ for the free evolution with flat initial state, and the superscript $\text{f}\to\text{f}$ for conditioning on flat initial and final states.

One finds the same scalings for the cumulants at short time in both cases, but with different rescaled cumulants. This is not surprising as one expects to have several universality classes at short time, similarly to what happens for KPZ universality on the infinite line \cite{C2011.1}. On the other hand, in the long time limit corresponding to the stationary state, one finds the same cumulants $\langle(\xi_{t}^{\text{f}\to\text{f}})^{k}\rangle_{\text{c}}\simeq\langle(\xi_{t}^{\text{f}\to})^{k}\rangle_{\text{c}}$, which are proportional to $t$ for large $t$. A better correspondence at finite $t$ is however obtained by decomposing the current fluctuations in the flat $\to$ flat case as $\xi_{t}^{\text{f}\to\text{f}}=\xi_{0\to t/2}+\xi_{t/2\to t}$ where $\xi_{0\to t/2}$ and  $\xi_{t/2\to t}$ represent the first and last half of the evolution. For large $t$, $\xi_{0\to t/2}$ and $\xi_{t/2\to t}$ are independent and have the same statistics as $\xi_{t/2}^{\text{f}\to}$. The additivity of cumulants for independent variables implies $\langle(\xi_{t}^{\text{f}\to\text{f}})^{k}\rangle_{\text{c}}\simeq2\langle(\xi_{t/2}^{\text{f}\to})^{k}\rangle_{\text{c}}$. Equivalently, $\langle(\xi_{t}^{\text{f}\to})^{k}\rangle_{\text{c}}/t^{k-2/3}\simeq2^{k-\frac{5}{3}}\langle(\xi_{2t}^{\text{f}\to\text{f}})^{k}\rangle_{\text{c}}/t^{k-2/3}$ holds at large $t$, with rather good agreement for moderately large $t$, and very good agreement for the variance $k=2$ at small $t$ too, see figure \ref{Fig cumulants flat stat}.
\begin{figure}
  \begin{center}
    \begin{tabular}{c}\includegraphics[width=100mm]{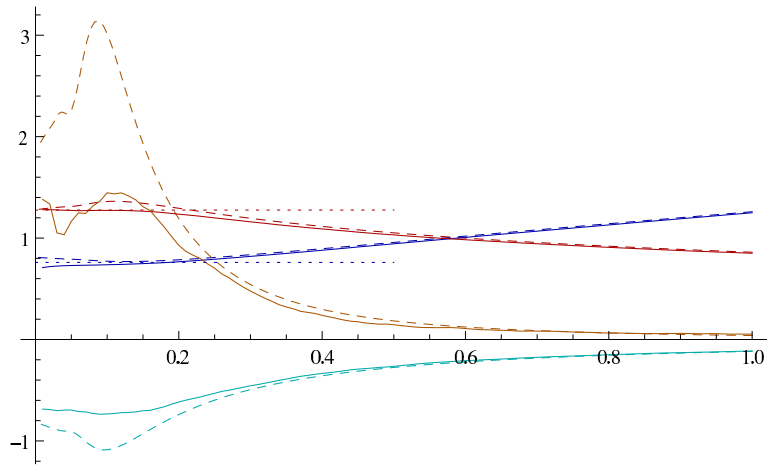}\end{tabular}
  \end{center}
  \caption{Rescaled first cumulants of the current fluctuations $\xi_{t}$ plotted as a function of $t$. The solid lines correspond to $\langle\xi_{t}^{k}\rangle_{\text{c}}/t^{k-2/3}$ for an evolution starting on the flat configuration and not conditioned on the final state, and are the result of simulations with $N=1000$ particles on $L=2000$ sites averaged over $326656$ independent realizations. The dashed lines correspond to $2^{k-5/3}\langle\xi_{2t}^{k}\rangle_{\text{c}}/(2t)^{k-2/3}$ for an evolution conditioned on flat initial and final states, computed from the generating function (\ref{G flat flat}). From top to bottom on the right side of the graph, they represent the rescaled average (blue), variance (red), fourth cumulant (orange) and third cumulant (cyan). The lower horizontal dotted line is $-2^{-2/3}$ times the mean value of GOE Tracy-Widom distribution. The upper horizontal dotted line is $2^{-1/3}$ times the variance of GOE Tracy-Widom distribution.}
  \label{Fig cumulants flat stat}
\end{figure}

Apart from the total current $Q$, another interesting quantity is the (local, time-integrated) current $Q_{i}$ between sites $i$ and $i+1$ (at half-filling $\overline{\rho}=1/2$ only because of the necessity to consider a moving reference frame with velocity $1-2\,\overline{\rho}$ in order to see fluctuations characteristic of KPZ universality). When the evolution is conditioned on both the initial and final configuration, $Q$ and $Q_{i}$ are closely related since particles can not overtake each other (this can also be understood by a simple similarity transformation of the deformed Markov matrix $M(\gamma)$ with a diagonal change of basis in the configuration basis). In particular, if the initial and the final configurations are identical, one has $Q=LQ_{i}$ at the final time $T$. This is not the case any more for an evolution conditioned only on the initial configuration.

On the infinite line $\mathbb{Z}$, the statistics of $Q_{i}$ have been investigated in much detail \cite{C2011.1} for an evolution not conditioned on the final state. For flat initial condition, the probability density of $-2^{2/3}\,\xi_{t}^{(i)}/t^{1/3}$ with $\xi_{t}^{(i)}=2(Q_{i}^{\text{f}\to}-JT)/\sqrt{L}$ and (\ref{scaling T}) is given \cite{S2005.1,BFPS2007.1} by the (derivative of the) GOE Tracy-Widom distribution from random matrix theory. This result presumably also holds at any rescaled time $\tau=T/L$ on the Euler time scale $T\sim L$, and in the limit $t\to0$ on the KPZ time scale. If the initial condition has particles only at odd sites, we observe rather strong finite size corrections for the mean value at small $t$ of $\xi_{t}^{(i)}/t^{1/3}$ with $i$ even, which disappear for $i$ odd, see figure \ref{Fig cumulants Qi}.
\begin{figure}
  \begin{center}
    \begin{tabular}{ll}
      \begin{tabular}{c}\includegraphics[width=70mm]{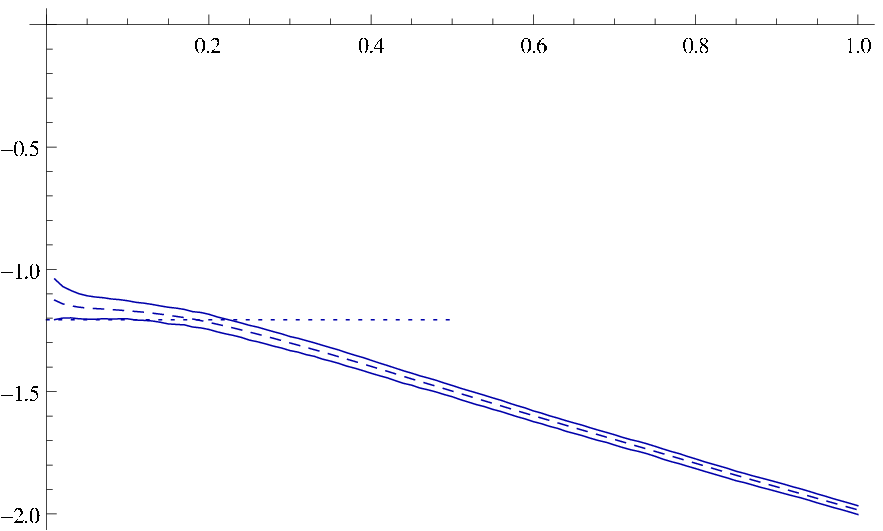}\end{tabular}
      &
      \begin{tabular}{c}\includegraphics[width=70mm]{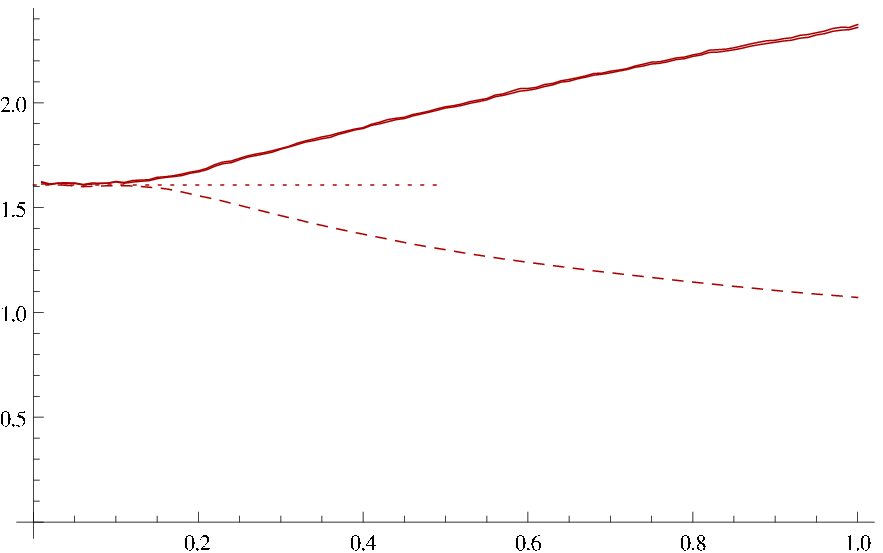}\end{tabular}
    \end{tabular}
  \end{center}
  \caption{Mean value (left) and variance (right) of current fluctuations for periodic TASEP with flat initial condition and no conditioning on the final state, plotted in terms of the rescaled time $t$ (\ref{scaling T}). The solid lines, corresponding to the mean value and variance of the fluctuations $-2^{2/3}\xi_{t}^{(i)}/t^{1/3}$ of the current $Q_{i}$ between sites $i$ and $i+1$, are the results of simulations with $N=1000$ particles on $L=2000$ sites, averaged over $134144$ realizations for an initial condition with particles only on odd sites. On the left, the upper solid line, which shows large finite size corrections for small $t$, corresponds to $i=0$ while the lower solid line, which does not show any noticeable finite size effect, corresponds to $i=1$; on the right, the two solid lines are almost superposed. The dashed lines correspond to the fluctuations $\xi_{t}$ of the total current $Q$. On the left, the mean value $-2^{2/3}\langle\xi_{t}\rangle/t^{1/3}$ is plotted, and on the right the rescaled variance $2^{1/3}t^{-4/3}\langle\xi_{t}^{2}\rangle_{\text{c}}$. The dotted lines correspond to the mean value (left) and the variance (right) of GOE Tracy-Widom distribution.}
  \label{Fig cumulants Qi}
\end{figure}

The precise relation between the GOE Tracy-Widom distribution for the fluctuations of $Q_{i}^{\text{f}\to}$ and the large deviations of the total current is currently not known. However, since $Q^{\text{f}\to}=\sum_{i=1}^{L}Q_{i}^{\text{f}\to}$, their mean values must be equal. More precisely, for an initial condition with particles only at odd sites, the mean value of $\xi_{t}$ is equal to $(\langle\xi_{t}^{(0)}\rangle+\langle\xi_{t}^{(1)}\rangle)/2$ for a finite system. The finite-size corrections to $\xi_{t}^{(0)}$ for small $t$ are responsible for the not so good convergence to the mean value of GOE Tracy-Widom in figure \ref{Fig cumulants flat stat}, see also figure \ref{Fig cumulants Qi}. We have no explanation however for the numerical coincidences for the variance at small time
\begin{eqnarray}
&&\fl\hspace{5mm} 1.607781034581 \approx \mathrm{Var}_{\text{GOE}}
=\lim_{t\to0}2^{4/3}t^{-2/3}\langle(\xi_{t}^{(i),\text{f}\to})^{2}\rangle_{\text{c}}\nonumber\\
&&\fl\hspace{52mm} \approx \lim_{t\to0}2^{1/3}t^{-4/3}\langle(\xi_{t}^{\text{f}\to})^{2}\rangle_{\text{c}}\\
&&\fl\hspace{52mm} \approx \lim_{t\to0}2^{2/3}t^{-4/3}\langle(\xi_{t}^{\text{f}\to\text{f}})^{2}\rangle_{\text{c}}
\approx 1.6074631729182734577\;,\nonumber
\end{eqnarray}
with $\mathrm{Var}_{\text{GOE}}$ the variance of GOE Tracy-Widom distribution, see (\ref{cumulants small t}), figure \ref{Fig cumulants flat stat} and figure \ref{Fig cumulants Qi}.
\end{subsection}
\end{section}

\begin{section}{Conclusions}
Current fluctuations for periodic TASEP on the relaxation scale are studied in this paper using large system size asymptotics of eigenvalues and eigenvectors of the generator of the evolution. For technical reasons, our results are restricted to evolutions conditioned on simple initial and final states. An exact formula for the generating function of current fluctuations is obtained as a sum over eigenstates. In the special case of flat initial and final configurations, it leads to a simple expression (\ref{Xi[int,Ai]}) for the large deviations of the current at the early stages of the relaxation, written in terms of the Airy function.

Extending these results to more general initial and final states would be interesting in order to fully describe the process on the relaxation scale. Removing the conditioning over the final state would also allow to understand better the relation with the Tracy-Widom distributions that describe current fluctuations on the infinite line, by finding short time large deviation functions for the total current taking their minimum at the mean value of the corresponding Tracy-Widom distribution.

The results obtained in this paper should presumably extend to all models in one-dimensional KPZ universality. It would be interesting to recover them directly from stochastic Burgers' / KPZ equation with periodic boundary conditions, using the replica method with precise asymptotics for the attractive $\delta$-Bose gas in finite volume.

\noindent\textbf{Acknowledgements:} It is a pleasure to thank C.~Bahadoran, M.~Bauer, K.~Mallick and H.~Spohn for useful discussions.
\end{section}

\appendix
\begin{section}{Total integrated current for periodic Burgers' equation}
\label{appendix Burgers}
In this appendix, we check the expression (\ref{R}) for the first correction at large times to the total current in the inviscid Burgers' equation (\ref{Burgers}) by working out a few simple examples with piecewise linear initial condition $\rho_{0}$, for which the solution of the partial differential equation can be computed easily: the linear parts evolve in time as $\half-\frac{x-x_{0}}{\tau-\tau_{0}}$, and a discontinuity at $x_{0}$ with densities respectively $\rho_{-}$ and $\rho_{+}$ on the left and on the right leads if $\rho_{-}<\rho_{+}$ to a shock that moves with a velocity $1-\rho_{+}(\tau)-\rho_{-}(\tau)$ imposed by conservation of density, and if $\rho_{-}>\rho_{+}$ to a rarefaction fan with density profile $\rho(x,\tau)=\half-\frac{x-x_{0}}{2\tau}$ in the interval $x_{0}+(1-2\rho_{-})\tau<x<x_{0}+(1-2\rho_{+})\tau$ until one of the extremities is absorbed by a shock.

\begin{subsection}{Step initial condition}
The step initial condition is defined as the periodic profile with periodicity $1$ equal for $x\in[0,1)$ to $\rho_{0}(x)=\rho_{+}\openone_{\{x\in[0,a)\}}+\rho_{-}\openone_{\{x\in[a,1)\}}
$, with $0\leq\rho_{-}<\rho_{+}\leq1$ and $0<a<1$. The corresponding average density is $\overline{\rho}=a\rho_{+}+(1-a)\rho_{-}$. We assume $a<\thalf$, the case $a>\thalf$ being accessible by changing $\rho_{0}$ to $1-\rho_{0}$. At the beginning of the evolution, the shock initially located at $z_{0}=0$ moves at the constant speed $1-\rho_{+}-\rho_{-}$, while a rarefaction fan opens at position $a$, with its left side moving at speed $1-2\rho_{+}$ and its right side at speed $1-2\rho_{-}$, see figure \ref{Fig Burgers Step Lin+}. This continues until time $\tau_{1}=a/(\rho_{+}-\rho_{-})$ when the interval of density $\rho_{+}$ disappears as the left of the fan merges with the shock. Then, the position of the shock moves as $z(\tau)=a-2\sqrt{a}\sqrt{\rho_{+}-\rho_{-}}\sqrt{\tau}+(1-2\rho_{-})\tau$ until time $\tau_{2}=(4a(\rho_{+}-\rho_{-}))^{-1}$ when the interval of density $\rho_{-}$ disappears as the right of the fan merges with the shock. Finally, after $\tau_{2}$, the shock moves as $z(\tau)=(1-2\,\overline{\rho})\tau+a-\thalf$. The corresponding density profile is then $\rho(x,\tau)=\half-\frac{x-a}{2\tau}$ for $z(\tau)<x<z(\tau)+1$. Equivalently, in a reference frame moving at velocity $1-2\,\overline{\rho}$, one has $\rho(x+(1-2\rho)\tau,\tau)=\overline{\rho}-\frac{x-a}{2\tau}$ for $a-\thalf<x<a+\thalf$. After some calculations, one finds for the total integrated current up to time $\tau>\tau_{2}$
\begin{equation}
\mathcal{Q}_{\tau}[\rho_{0}]=\overline{\rho}(1-\overline{\rho})\tau-\frac{(\overline{\rho}-\rho_{-})(\rho_{+}-\overline{\rho})}{2(\rho_{+}-\rho_{-})}+\frac{1}{48\tau}\;.
\end{equation}
The constant term agrees with the general expression (\ref{R}) for $\mathcal{R}[\rho_{0}]$ with $\kappa=a-1$ modulo $1$.
\begin{figure}
  \begin{center}
    \begin{tabular}{lll}
      \hspace{-3.5mm}
      \begin{tabular}{c}\includegraphics[width=70mm]{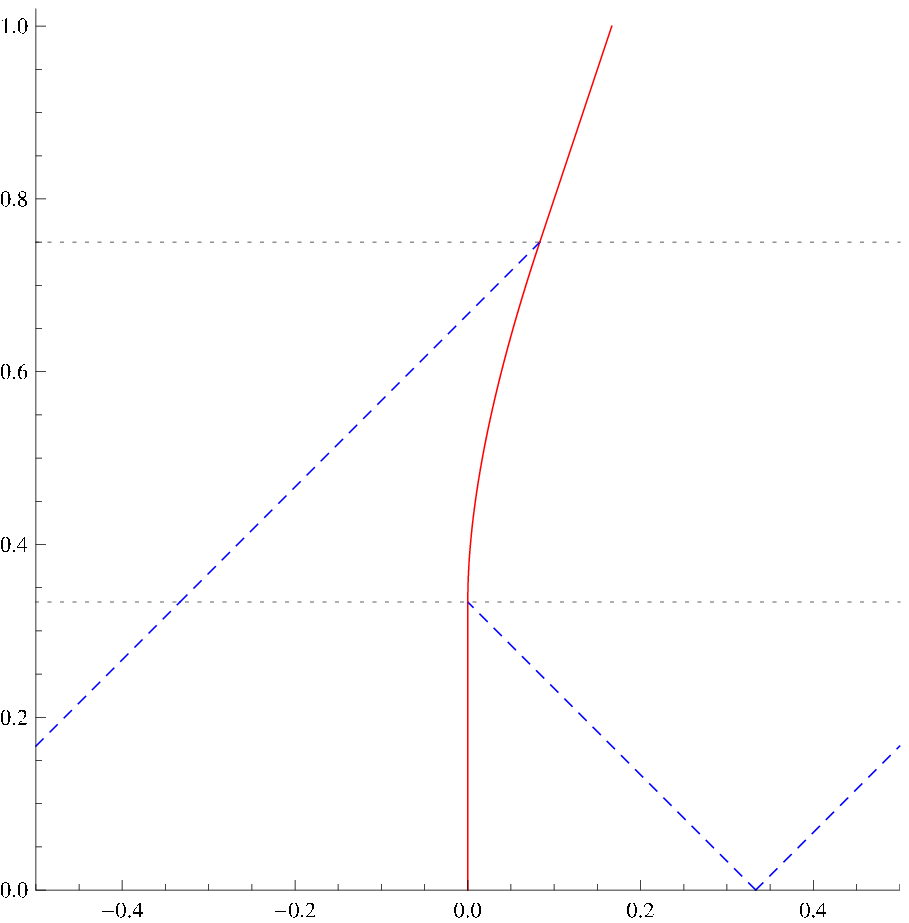}\end{tabular}
      \begin{picture}(0,0)
        \put(-50,-16){$0$}
        \put(-30,-26){$1$}
        \put(-62,20.5){rarefaction}
        \put(-55,15.5){fan}
        \put(-29,6){rarefaction}
        \put(-22,1){fan}
        \put(-75,-10){$\tau_{1}$}
        \put(-75,17.5){$\tau_{2}$}
      \end{picture}
      &&
      \begin{tabular}{c}\includegraphics[width=70mm]{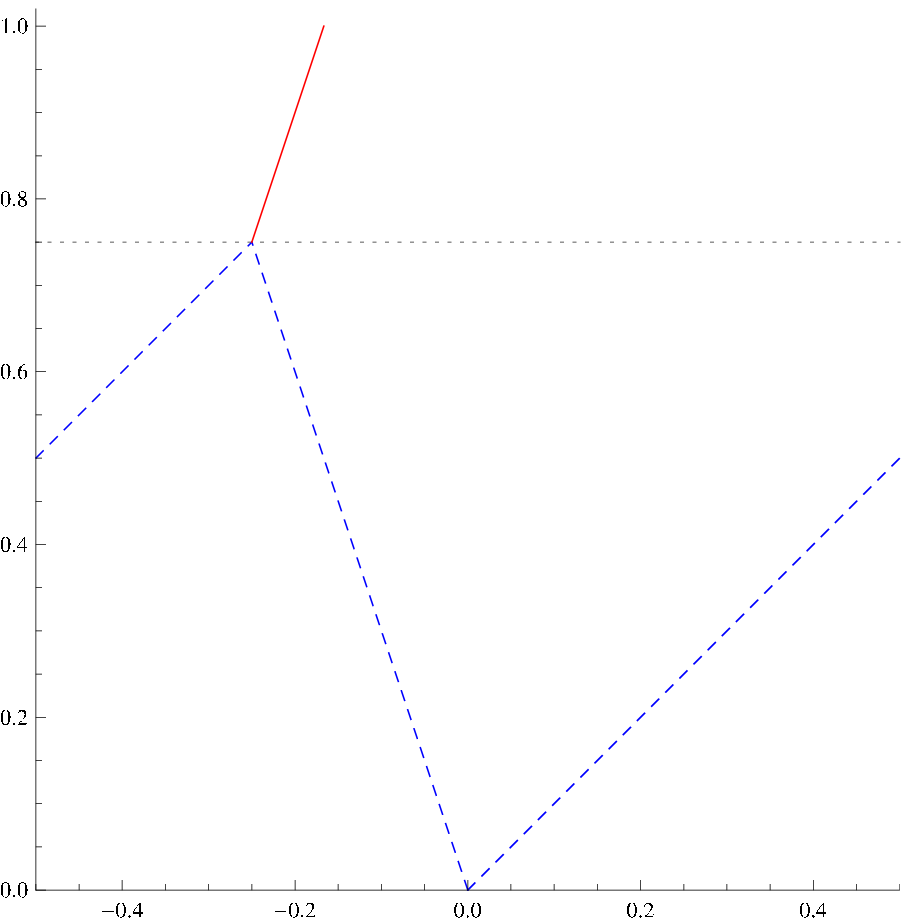}\end{tabular}
      \begin{picture}(0,0)
        \put(-62,-5){linear}
        \put(-67,-10){increasing}
        \put(-62.5,-15){profile}
        \put(-38,20.5){rarefaction}
        \put(-31,15.5){fan}
        \put(-75,17.5){$\tau_{1}$}
      \end{picture}
    \end{tabular}
  \end{center}
  \caption{Evolution of the density profile from Burgers' equation (\ref{Burgers}) with periodic boundary condition for unit step (left) and linear increasing (right) initial condition at average density $\overline{\rho}=1/3$. The horizontal axis correspond to space and the vertical axis to time. The solid, red curve represents the shock. Dashed, blue lines divide regions with a different behaviour for the density profile. Dotted horizontal line indicate the times at which edges of rarefaction fans merge with the shock.}
  \label{Fig Burgers Step Lin+}
\end{figure}
\end{subsection}

\begin{subsection}{Linear decreasing profile}
We consider the initial profile $\rho_{0}(x)=\overline{\rho}-\alpha(x-\thalf)$ for $0<x<1$ and $0<\alpha<2\min(\overline{\rho},1-\overline{\rho})$, with a shock initially located at $z_{0}=0$. At any time $\tau>0$, the suitable solution of (\ref{Burgers}) is equal in the moving frame to $\rho(x+(1-2\,\overline{\rho})\tau,\tau)=\overline{\rho}-\frac{x-\half}{2\tau+\alpha^{-1}}$ for $0<x<1$. This implies $\kappa=\thalf$ and a position of the shock $z(\tau)=(1-2\,\overline{\rho})\tau$. The calculation of the current from the previous expression for $\rho(x,\tau)$ gives
\begin{eqnarray}
&& \mathcal{Q}_{\tau}[\rho_{0}]=\overline{\rho}(1-\overline{\rho})\tau-\frac{\alpha^{2}\tau}{12+24\alpha\tau}\\
&&\hspace{9mm} \underset{\tau\to\infty}{\simeq}\overline{\rho}(1-\overline{\rho})\tau-\frac{\alpha}{24}+\frac{1}{48\tau}-\frac{1}{96\alpha\tau^{2}}+\frac{1}{192\alpha^{2}\tau^{3}}\;.\nonumber
\end{eqnarray}
The constant term matches with (\ref{R}).
\end{subsection}

\begin{subsection}{Linear increasing profile}
We consider the initial profile $\rho_{0}(x)=\overline{\rho}+\alpha(x-\thalf)$ for $0<x<1$ and $0<\alpha<2\min(\overline{\rho},1-\overline{\rho})$. In the beginning of the evolution, a rarefaction fan opens at position $0$, see figure \ref{Fig Burgers Step Lin+}. Until time $\tau_{1}=\frac{1}{2\alpha}$, the density profile is $\rho(x,\tau)=\half-\frac{x-\half-(1-2\,\overline{\rho})\tau_{1}}{2(\tau-\tau_{1})}$ for $x$ in the interval $[(1-2\rho+\alpha)\tau,(1-2\rho-\alpha)\tau+1]$ and  $\rho(x,\tau)=\half-\frac{x}{2\tau}$ for $x$ in the interval $[(1-2\rho-\alpha)\tau,(1-2\rho+\alpha)\tau]$. Then, at time $\tau_{1}$, the linearly increasing portion vanishes and a shock forms leading to the density profile $\rho(x+(1-2\,\overline{\rho})\tau,\tau)=\overline{\rho}-\frac{x}{2\tau}$ for $-\thalf<x<\thalf$. A space-time integration leads for $\tau>\tau_{1}$ to
\begin{equation}
\mathcal{Q}_{\tau}[\rho_{0}]=\overline{\rho}(1-\overline{\rho})\tau-\frac{\alpha}{12}+\frac{1}{48\tau}\;,
\end{equation}
which can be recovered directly from (\ref{Q Burgers' inviscid asymptotics}), (\ref{R}) with $\kappa=0$. This expression is however exact for $\tau>\tau_{1}$, unlike (\ref{Q Burgers' inviscid asymptotics}) which is the beginning of a large $\tau$ asymptotics.
\end{subsection}
\end{section}

\begin{section}{Bethe ansatz for TASEP}
\label{appendix Bethe}
In this appendix, we summarize some known results about Bethe ansatz for the first eigenstates of TASEP, in particular large $L$, $N$ asymptotics of normalization of eigenvectors and components of the eigenvectors corresponding to unit step density profile. We also derive the asymptotics (\ref{PhiF}) for the components of the eigenvectors corresponding to a flat density profile.

\begin{subsection}{Bethe equations and their solution}
From Bethe ansatz, each eigenstate of periodic TASEP for a finite system of length $L$ with $N$ particles is completely characterized by $N$ complex numbers $y_{j}$, $j=1,\ldots,N$, the \textit{Bethe roots}, that satisfy a set of $N$ polynomial equations called the \textit{Bethe equations}:
\begin{equation}
\label{Bethe equations}
\rme^{L\gamma}(1-y_{j})^{L}=(-1)^{N-1}\prod_{k=1}^{N}\frac{y_{j}}{y_{k}}\;.
\end{equation}
Multiplying both sides of the Bethe equations by $\rme^{-L\gamma}y_{j}^{-N}$ and taking the power $1/L$ gives \cite{PP2007.1,P2013.1}
\begin{equation}
\label{g(yj)}
g(y_{j})=\rme^{\frac{2\rmi\pi k_{j}}{L}-b}\;,
\end{equation}
where the $k_{j}$'s, distinct modulo $L$, are integers (half-integers) if $N$ is odd (even). The function $g$ is defined by
\begin{equation}
\label{g}
g(y)=\frac{1-y}{y^{\overline{\rho}}}\;,
\end{equation}
and $b$ is solution of
\begin{equation}
\label{b[y]}
b-\gamma=\frac{1}{L}\sum_{j=1}^{N}\log y_{j}\;.
\end{equation}
The branch cut of $g$ due to the non-integer power is taken as $\mathbb{R}^{-}$, which leads to the branch cuts $\rme^{\pm\rmi\pi\rho}\rho^{-\rho}(1-\rho)^{-1+\rho}[1,\infty)$ for the inverse function $g^{-1}$.

The equation (\ref{b[y]}) can be solved numerically for $b$ with high accuracy using Newton's method. At each step, the $y_{j}$'s are computed by inverting $g$ in (\ref{g(yj)}) using again Newton's method. It is possible to obtain very accurate expressions for the Bethe roots $y_{j}$, with several hundred significant digits. Such accurate values are needed in order to fully exploit the power of Richardson extrapolation for obtaining precise asymptotics of various quantities from a few finite size values, see table \ref{Table extrapolation} for an example.

The question of the completeness of Bethe ansatz for finite systems has not been fully solved yet, see however \cite{D1993.1,LSA1995.1,LSA1997.1,PP2007.1}. Nevertheless, one observes that the number of possible choices for the $k_{j}$'s with $0<k_{1}<\ldots<k_{N}\leq L$ is equal to the number of configurations $\Omega=\C{L}{N}$. Comparison with exact diagonalization for systems up to size $L=18$ seems to indicate that each such choice of the $k_{j}$'s corresponds to an eigenstate of the Markov matrix with (\ref{b[y]}) having a unique finite solution, if $\Re\,\gamma$ is large enough so that the solution $b$ of (\ref{b[y]}) is such that the circle of center $0$ and radius $\rme^{-b}$ does not cross the branch cuts of $g^{-1}$. Numerics indicate that $\Re\,\gamma\geq0$ is sufficient for all the eigenstates. The stationary state $k_{j}^{0}=j-(N+1)/2$ with $\gamma$ close to $0$ is special since the circle crosses the branch cuts of $g^{-1}$, which is however not a problem since the points $\rme^{2\rmi\pi k_{j}^{0}/L-b}$ stay on the same side of the branch cuts in that case; this special case leads to the singular solution $c_{0}(s)\to-\infty$ when $s\to0$ of (\ref{c(s)}) in the large $L$ limit.

An alternative approach was used in \cite{GM2004.1,GM2005.1} to characterize the solutions of the Bethe equations (\ref{Bethe equations}) at $\gamma=0$, by rewriting $g(y_{j})^{L}=\rme^{-bL}$ as $P((1+y_{j})/(1-y_{j}))$ with $P$ a polynomial of degree $L$ with coefficients depending on $b$. Using a particular labelling of the $L$ roots of $P$, the eigenstates were then identified as choices of $N$ distinct roots of $P$ among $L$. As noted in \cite{MSS2012.1}, however, this identification fails for some eigenstates of large enough systems. This is due to the fact that with this specific labelling for the roots of $P$, changing the imaginary part of $b$ can induce a cyclic relabelling of the roots of $P$.
\end{subsection}

\begin{subsection}{Eigenvalues and eigenvectors}
The eigenvalue of $M(\gamma)$ corresponding to a given solution of the Bethe equations (\ref{Bethe equations}) is given by
\begin{equation}
E_{r}(\gamma)=\sum_{j=1}^{N}\frac{y_{j}}{1-y_{j}}\;.
\end{equation}
The corresponding eigenvalue for the translation operator $U$, $U|x_{1},\ldots,x_{N}\rangle=|1+x_{1},\ldots,1+x_{N}\rangle$, is equal to
\begin{equation}
\rme^{2\rmi\pi p_{r}/L}=\rme^{N\gamma}\prod_{j=1}^{N}(1-y_{j})=\prod_{j=1}^{N}\rme^{2\rmi\pi k_{j}/L}\;.
\end{equation}

The components of the left and right eigenvectors of $M(\gamma)$ with particles at positions $x_{j}$, $j=1,\ldots,N$ with $1\leq x_{1}<\ldots<x_{N}\leq L$ are given by the \textit{Bethe ansatz} as linear combinations of all $N!$ permutations of $N$ plane waves with pseudo-momenta $\gamma+\log(1-y_{k})$. For TASEP, the sum over permutations reduces to a determinant. One has
\begin{eqnarray}
\label{psiR[y]}
&&\fl\hspace{10mm} \langle\vec{x}|\psi_{r}\rangle=(-\rmi)^{\frac{N(N-1)}{2}}N^{-N/2}\Bigg(\prod_{j=1}^{N}y_{j}^{\frac{N+1}{2}}\Bigg)\det\Big(y_{k}^{-j}(1-y_{k})^{x_{j}}\rme^{\gamma x_{j}}\Big)_{j,k=1,\ldots,N}\\
\label{psiL[y]}
&&\fl\hspace{10mm} \langle\psi_{r}|\vec{x}\rangle=\rmi^{\frac{N(N-1)}{2}}\rme^{\frac{2\rmi\pi p_{r}}{L}}N^{-N/2}\Bigg(\prod_{j=1}^{N}y_{j}^{-\frac{N+1}{2}}\Bigg)\det\Big(y_{k}^{j}(1-y_{k})^{-x_{j}}\rme^{-\gamma x_{j}}\Big)_{j,k=1,\ldots,N}\;.
\end{eqnarray}
These Bethe eigenstates are not normalized. The factors in front of the determinants are chosen in prevision for the thermodynamic limit.

For any configuration $\vec{x}$, the reversed configuration $\vec{\tilde{x}}$ is defined by $\tilde{x}_{j}=L+1-x_{N+1-j}$. One has the symmetry relation
\begin{equation}
\label{psiL[psiR]}
\langle\psi|\vec{x}\rangle=\langle\vec{\tilde{x}}|\psi\rangle\;,
\end{equation}
which is a consequence of the fact that transposing the evolution operator of TASEP is equivalent to reversing space.
\end{subsection}

\begin{subsection}{Normalization of Bethe eigenstates}
The norm of Bethe eigenstates is in general given by the Gaudin determinant \cite{GMCW1981.1,K1982.1}
It reduces for TASEP to the explicit expression \cite{MSS2012.2,P2015.2}
\begin{equation}
\langle\psi_{r}|\psi_{r}\rangle=\rme^{\frac{2\rmi\pi p_{r}}{L}}\Bigg(\frac{1}{N}\sum_{j=1}^{N}\frac{y_{j}}{\overline{\rho}+(1-\overline{\rho})y_{j}}\Bigg)\Bigg(\prod_{j=1}^{N}\Big(\frac{1-\overline{\rho}}{\overline{\rho}}+y_{j}^{-1}\Big)\Bigg)\;,
\end{equation}
whose asymptotics can be obtained using the Euler-Maclaurin formula. With $\Omega=\C{L}{N}$ the total number of configurations, one has \cite{P2015.2}
\begin{equation}
\label{norm psi}
\fl\hspace{20mm}
\frac{\Omega}{\langle\psi_{r}|\psi_{r}\rangle}\simeq\frac{(2\pi)^{-1/2}\,\rme^{2\pi c_{r}}}{\sqrt{1+\rme^{2\pi c_{r}}}\,\chi_{r}''(2\pi c_{r})}\,\frac{(\prod_{a\in A_{0}^{+}}\sqrt{c_{r}+\rmi a})(\prod_{a\in A_{0}^{-}}\sqrt{c_{r}-\rmi a})}{(\prod_{a\in A^{+}}\sqrt{c_{r}-\rmi a})(\prod_{a\in A^{-}}\sqrt{c_{r}+\rmi a})}\;,
\end{equation}
where $\chi_{r}$ is defined in (\ref{chi[A,zeta]}) and $c_{r}$ is the solution of (\ref{c(s)}). This leads to (\ref{norm phi}) after changing the normalization of the eigenvectors as $\langle\phi_{r}|=\lambda_{r}^{-1}\langle\psi_{r}|$ and $|\phi_{r}\rangle=\lambda_{r}^{-1}|\psi_{r}\rangle$ with
\begin{eqnarray}
\label{lambda}
&& \lambda_{r}=(-\rmi)^{(m_{r}^{+}+m_{r}^{-})^{2}}\rme^{\rmi\pi\big((\sum_{a\in A_{0}^{+}}a)-(\sum_{a\in A_{0}^{-}}a)\big)}(1+\rme^{2\pi c_{r}})^{1/4}\nonumber\\
&&\hspace{10mm} \times\frac{\big(\prod_{a\in A^{+}}(c_{r}-\rmi a)^{1/4}\big)\big(\prod_{a\in A^{-}}(c_{r}+\rmi a)^{1/4}\big)}{\big(\prod_{a\in A_{0}^{+}}(c_{r}+\rmi a)^{1/4}\big)\big(\prod_{a\in A_{0}^{-}}(c_{r}-\rmi a)^{1/4}\big)}\;.
\end{eqnarray}
\end{subsection}

\begin{subsection}{Flat configuration (\texorpdfstring{$1/\overline{\rho}$}{1/(rho bar)} integer)}
We consider the flat configuration $\mathcal{F}_{X}$ with particles at positions $x_{j}=X+(j-1)/\overline{\rho}$, $j=1,\ldots,N$ and $\overline{\rho}^{-1}$ integer. The determinants in (\ref{psiR[y]}) and (\ref{psiL[y]}) are then Vandermonde determinants:
\begin{eqnarray}
&& \langle\mathcal{F}_{X}|\psi\rangle=(-\rmi)^{\frac{N(N-1)}{2}}\rme^{\frac{2\rmi\pi p_{r}X}{L}}\rme^{\frac{(N-1)L\gamma}{2}}N^{-N/2}\Big(\prod_{j=1}^{N}y_{j}^{\frac{N-1}{2}}\Big)\\
&&\hspace{47mm} \times\!\!\!\prod_{1\leq j<\ell\leq N}\!\!\!\Big(g(y_{\ell})^{1/\overline{\rho}}-g(y_{j})^{1/\overline{\rho}}\Big)\;,\nonumber
\end{eqnarray}
where the function $g$ is defined by (\ref{g}). This expression can be simplified further by noting that the Bethe equations precisely give an explicit expression (\ref{g(yj)}) for $g(y_{j})$ in terms of the (half-)integers $k_{j}$. Using (\ref{b[y]}) to simplify the single product of the $y_{j}$'s, one has
\begin{equation}
\langle\mathcal{F}_{X}|\psi\rangle=(-\rmi)^{\frac{N(N-1)}{2}}\rme^{\frac{2\rmi\pi p_{r}X}{L}}N^{-N/2}\!\!\!\prod_{1\leq j<\ell\leq N}\!\!\!\big(\rme^{2\rmi\pi k_{\ell}/N}-\rme^{2\rmi\pi k_{j}/N}\big)\;.
\end{equation}
From the symmetry relation (\ref{psiL[psiR]}), the left eigenstate is given by $\langle\psi|\mathcal{F}_{X}\rangle=\langle\mathcal{F}_{L+1-X}|\psi\rangle$. We observe that these expressions are non-zero if and only if the $k_{j}$'s are all distinct modulo $N$. For the first eigenstates, described in figure \ref{Fig sets A}, it is equivalent to the constraints $A_{0}^{+}=A^{-}$ and $A_{0}^{-}=A^{+}$, which imply that all the sets have the same cardinal $m_{r}=m_{r}^{+}=m_{r}^{-}$ and that the total momentum $p_{r}=0$. Splitting the contributions to the double product coming from the Fermi sea $k_{j}^{0}=j-(N+1)/2$ and from the sets $A_{0}^{\pm}$, $A^{\pm}$, one finds
\begin{equation}
\fl \prod_{1\leq j<\ell\leq N}\!\!\!\big(\rme^{2\rmi\pi k_{\ell}/N}-\rme^{2\rmi\pi k_{j}/N}\big)=(-1)^{\big(\sum_{a\in A_{0}^{+}}a\big)+\big(\sum_{a\in A_{0}^{-}}a\big)}\!\!\!\prod_{1\leq j<\ell\leq N}\!\!\!\big(\rme^{2\rmi\pi k_{\ell}^{0}/N}-\rme^{2\rmi\pi k_{j}^{0}/N}\big)\;.
\end{equation}
The double product reduces to a simple product by factoring out $\rme^{2\rmi\pi k_{\ell}^{0}/N}$ and making the change of variables $j\to j+\ell$. The remaining simple product can be computed using the symmetry $j\leftrightarrow N-j$. One finds
\begin{equation}
\prod_{1\leq j<\ell\leq N}\!\!\!\big(\rme^{2\rmi\pi k_{\ell}^{0}/N}-\rme^{2\rmi\pi k_{j}^{0}/N}\big)=\rmi^{\frac{N(N-1)}{2}}N^{N/2}\;.
\end{equation}
It finally gives the asymptotics $\langle\mathcal{F}_{X}|\psi\rangle\simeq\langle\psi|\mathcal{F}_{X}\rangle\simeq\Psi_{r}^{\mathcal{F}}$ with $\Psi_{r}^{\mathcal{F}}$, independent of $X$, given by
\begin{equation}
\label{PsiF}
\Psi_{r}^{\mathcal{F}}
=(-1)^{\big(\sum_{a\in A_{0}^{+}}a\big)+\big(\sum_{a\in A_{0}^{-}}a\big)}
\openone_{\{A_{0}^{+}=A^{-}\}}
\openone_{\{A_{0}^{-}=A^{+}\}}\;,
\end{equation}
which leads to (\ref{PhiF}) after the change of normalization above (\ref{lambda}).
\end{subsection}

\begin{subsection}{Step configuration}
We consider the step configuration $\mathcal{S}_{X}$ with particles at positions $x_{j}=X+j-1$. The corresponding component of the eigenvectors are
\begin{equation}
\fl \langle\mathcal{S}_{X}|\psi\rangle=(-\rmi)^{\frac{N(N-1)}{2}}\rme^{\frac{2\rmi\pi p_{r}X}{L}}\rme^{\frac{N(N-1)\gamma}{2}}N^{-N/2}\Big(\prod_{j=1}^{N}y_{j}^{-\frac{N-1}{2}}\Big)\prod_{j=1}^{N}\prod_{k=j+1}^{N}(y_{j}-y_{k})\;,
\end{equation}
and $\langle\psi|\mathcal{S}_{X}\rangle=\langle\mathcal{S}_{L-N+2-X}|\psi\rangle$. The large $L$ asymptotics of this expression was studied in \cite{P2015.2} using two-dimensional Euler-Maclaurin formula with various logarithmic and square root singularities at the borders of the summation range. It has an expansion in powers of $1/\sqrt{L}$ instead of $1/L$ for the flat case. One has
\begin{equation}
\langle\mathcal{S}_{X}|\psi\rangle\simeq\rme^{2\rmi\pi p_{r}\overline{\rho}}\langle\psi|\mathcal{S}_{-X}\rangle\simeq\rme^{\frac{2\rmi\pi p_{r}X}{L}}\rme^{-\frac{\sqrt{\overline{\rho}(1-\overline{\rho})}s\sqrt{L}}{2}}\,\Psi_{r}^{\mathcal{S}}\;
\end{equation}
with
\begin{eqnarray}
\label{PsiS}
&& \Psi_{r}^{\mathcal{S}}
=\frac{(\pi/2)^{m_{r}^{2}}}{(2\pi)^{m_{r}}}
\rme^{\rmi\pi\big((\sum_{a\in A_{0}^{+}}a)-(\sum_{a\in A_{0}^{-}}a)\big)}\\
&&\hspace{10mm} \times\omega(A_{0}^{+})\omega(A_{0}^{-})\omega(A^{+})\omega(A^{-})\omega(A_{0}^{+},A_{0}^{-})\omega(A^{+},A^{-})\nonumber\\
&&\hspace{10mm} \times(1+\rme^{2\pi c_{r}})^{1/4}
\frac{\big(\prod_{a\in A^{+}}(c_{r}-\rmi a)^{1/4}\big)\big(\prod_{a\in A^{-}}(c_{r}+\rmi a)^{1/4}\big)}{\big(\prod_{a\in A_{0}^{+}}(c_{r}+\rmi a)^{1/4}\big)\big(\prod_{a\in A_{0}^{-}}(c_{r}-\rmi a)^{1/4}\big)}\nonumber\\
&&\hspace{10mm} \times\exp\Big(\lim_{\Lambda\to\infty}-m_{r}^{2}\log\Lambda+\int_{-\Lambda}^{2\pi c_{r}}\rmd u\,\frac{(\chi_{r}''(u))^{2}}{2}\Big)\;.\nonumber
\end{eqnarray}
The combinatorial factors $\omega$ are defined in (\ref{omega(A)}). This leads to (\ref{PhiS}) after the change of normalization above (\ref{lambda}) that cancels the third line and some factors in the first line of (\ref{PsiS}).
\end{subsection}

\end{section}

\begin{section}{Saddle point analysis of the flat \texorpdfstring{$\to$}{->} flat case at short time}
\label{appendix saddle point}
In this appendix, we derive the expression (\ref{Xi[int,Ai]}), (\ref{g0[Xi]}) for the large deviation function of the current at short time for an evolution conditioned on flat initial and final configurations.

We start from the exact formula (\ref{P flat flat}) for the probability density of $\xi_{t}$ at arbitrary rescaled time $t$, and consider instead the probability density of $t^{-1/3}\xi_{t}$. Making the change of variables $c=t^{-2/3}d$ and using
\begin{equation}
\chi_{0}(x)\simeq\Bigg\{
\begin{array}{lll}
  \frac{(2x)^{5/2}}{15\pi}+\mathcal{O}(\sqrt{x}) && \Re\,x>0\\
  \frac{\rme^{x}}{\sqrt{2\pi}} && \Re\,x<0
\end{array}\;,
\end{equation}
the small $t$ limit of the integrand gives
\begin{equation}
\label{P[Z,h] flat flat asymptotics t0}
t^{1/3}P_{t}(t^{1/3}u)\simeq\frac{t^{-1/3}}{\mathcal{Z}_{t}}\int\frac{\rmd d}{\rmi\sqrt{2\pi}}\oint\frac{\rmd z}{2\rmi\pi z}\,\rme^{t^{-2/3}h(u,d,z)+\mathcal{O}(t^{2/3})}\;,
\end{equation}
with
\begin{eqnarray}
\label{h[int v]}
&& h(u,d,z)=\openone_{\{\Re\,d>0\}}\Big(\frac{(4\pi d)^{5/2}}{15\pi}-u\,\frac{(4\pi d)^{3/2}}{3\pi}+\pi d\Big)+\openone_{\{\Re\,d<0\}}2\pi d\nonumber\\
&&\hspace{21mm} +\int_{0}^{\infty}\rmd v\,\log\Big(1+z^{-1}\rme^{-\frac{16\pi^{3/2}}{3}\sqrt{-\rmi}(v+\rmi d)^{3/2}+4u\sqrt{\pi}\sqrt{\rmi}\sqrt{v+\rmi d}}\Big)\\
&&\hspace{21mm} +\int_{0}^{\infty}\rmd v\,\log\Big(1+z\,\rme^{-\frac{16\pi^{3/2}}{3}\sqrt{\rmi}(v-\rmi d)^{3/2}+4u\sqrt{\pi}\sqrt{-\rmi}\sqrt{v-\rmi d}}\Big)\;.\nonumber
\end{eqnarray}
The branch cuts of the square roots are chosen equal to $\mathbb{R}^{-}$ so that $\sqrt{v\pm\rmi d}$ is analytic in both half-planes $\Re\,d>0$ and $\Re\,d<0$. Because of the branch cuts $\rmi[\half,\infty)$ and $-\rmi[\half,\infty)$ in the variable $c$, the two half-planes become independent when $t\to0$.

It is convenient to make the change of variables $w=2^{1/3}\sqrt{\rmi}\sqrt{4\pi}\sqrt{v+\rmi d}$ and $w=2^{1/3}\sqrt{-\rmi}\sqrt{4\pi}\sqrt{v-\rmi d}$ respectively in the first and the second integral. This leads to
\begin{eqnarray}
\label{h[int w]}
&&\fl\hspace{5mm} h(u,d,z)=\openone_{\{\Re\,d>0\}}\Big(\frac{(4\pi d)^{5/2}}{20\pi}-u\,\frac{(4\pi d)^{3/2}}{3\pi}+\pi d\Big)+\openone_{\{\Re\,d<0\}}2\pi d\\
&&\fl\hspace{15mm}
+\int_{\Gamma_{+}}\frac{w\,\rmd w}{2^{5/3}\rmi\pi}\,\log\Big(1+z^{-1}\rme^{\frac{w^{3}}{3}+2^{2/3}uw}\Big)
+\int_{\Gamma_{-}}\frac{w\,\rmd w}{2^{5/3}\rmi\pi}\,\log\Big(1+z\,\rme^{\frac{w^{3}}{3}+2^{2/3}uw}\Big)\;.\nonumber
\end{eqnarray}
The contour $\Gamma_{+}$ is oriented from $\alpha_{+}=2^{1/3}\sqrt{\rmi}\sqrt{4\pi}\sqrt{\rmi d}$ to $\sqrt{\rmi}\,\infty$, while the contour $\Gamma_{-}$ is oriented from $\sqrt{-\rmi}\,\infty$ to $\alpha_{-}=2^{1/3}\sqrt{-\rmi}\sqrt{4\pi}\sqrt{-\rmi d}$.

In \ref{appendix saddle point u>0} and \ref{appendix saddle point u<0}, we show that, for $d\in\mathcal{D}_{u}$ with $\mathcal{D}_{u}$ some specific unbounded domain of the complex plane, the relation $\partial_{d}h(u,d,z)=\rmi\log(-z)$ holds, and the quantity $h(u,d,z)$ is thus affine in $d$. The saddle point equation for the variable $d$ then gives $z=-1$. Expanding at second order in $z$, one has
\begin{equation}
\label{h[Xi,d*,A]}
\fl h(u,d,z)\simeq\Xi(u)-\rmi(d-d_{*}(u))(z+1)+(A(u)-\rmi(d-d_{*}(u)))\frac{(z+1)^{2}}{2}+\mathcal{O}(z+1)^{3}\;,
\end{equation}
with $\Xi(u)=h(u,d,-1)$ independent of $d$, and some functions $d_{*}(u)$ and $A(u)$. In order to perform the Gaussian integration around the saddle point, we write $z=-1+\sigma\rmi y$ and $d=d_{*}(u)+\nu\rmi x$, $\sigma=\pm1$, $\nu=\pm1$ (it does not matter any more at this point whether $d_{*}(u)$ still belongs to the domain $\mathcal{D}_{u}$ since, $h(u,d,z)$ being affine in $d$, it can be continued analytically to all $d\in\mathbb{C}$ before moving the contour for $d$ through the saddle point $d_{*}(u)$). The integration over $y$ gives
\begin{equation}
t^{1/3}P_{t}(t^{1/3}u)\simeq-\sigma\nu\,\rme^{t^{-2/3}\Xi(u)}\frac{1}{\mathcal{Z}_{t}}\int\frac{\rmd x}{2\pi}\,\frac{\exp(-\frac{t^{-2/3}x^{2}}{2(A(u)+\nu x)})}{\sqrt{A(u)+\nu x}}\;.
\end{equation}
The saddle point for $x$ is $x=0$ and one can then replace $A(u)+\nu x$ by $A(u)$ in the integrand. After Gaussian integration in $u$, one finds (\ref{P[Z] flat flat asymptotics t0}) with $\Xi(u)=h(u,d,-1)$. The signs have to be chosen as $\sigma\nu=-1$.
\begin{figure}
  \begin{center}
    \begin{tabular}{c}\includegraphics[width=60mm]{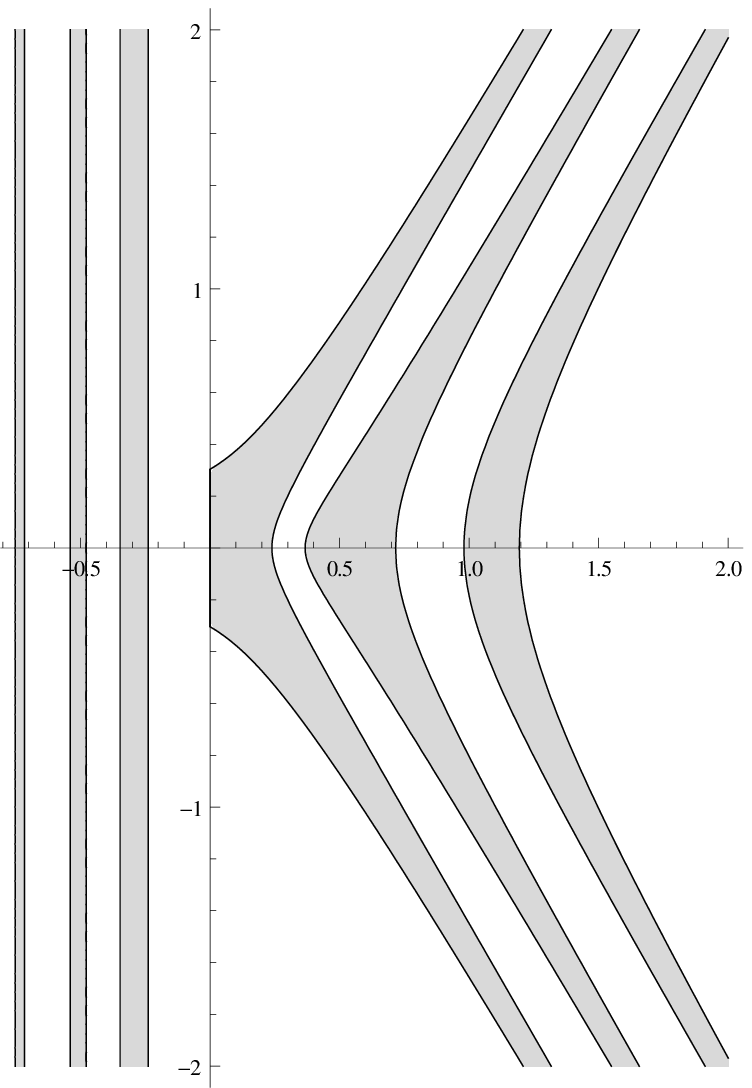}\end{tabular}
  \end{center}
  \caption{Domain $\mathcal{D}_{u}$ defined in (\ref{Du Red>0}) for $u>0$ and in (\ref{Du Red<0}) for $u<0$, plotted from left to right for $u=-3$, $u=-2$, $u=-1$, $u=1$, $u=3$ and $u=5$.}
  \label{Fig domain D}
\end{figure}

For technical reasons related to the complicated branch cut structure of the integrands in (\ref{h[int w]}) as a function of $w$, see figure \ref{Fig cuts w}, the choice of a suitable domain $\mathcal{D}_{u}$ for $d$ depends on the sign of $u$. When $u>0$, we will need to consider a domain $\mathcal{D}_{u}$ included in the half-plane $\Re\,d>0$, while for $u<0$, the domain $\mathcal{D}_{u}$ will be included in the half-plane $\Re\,d<0$. As shown on figure \ref{Fig c}, it is always possible to deform the contours $c_{r}(\rmi s)$, $s\in\mathbb{R}$ in the finite time expression (\ref{P[G]}), (\ref{G flat flat}) to a unique contour with $\Re\,c>0$, which makes the saddle point analysis straightforward when $u>0$. Deforming the contours $c_{r}(\rmi s)$, $s\in\mathbb{R}$ to a contour with $\Re\,c<0$ is on the other hand not possible. One can however always make some portion of the contours pass through the hole $\rmi[-\half,\half]$ between the branch cuts in figure \ref{Fig c}. The hole then closes after making the change of variable from $c$ to $d$ and taking the limit $t\to0$, and the integral over $d$ in (\ref{P[Z,h] flat flat asymptotics t0}) can be decomposed as an integral with $\Re\,d>0$ plus an integral with $\Re\,d<0$. Since only the part with $\Re\,d<0$ seems to possess a proper saddle point, we assume that the contribution of the integral with $\Re\,d>0$ is negligible in the small $t$ limit when $u<0$. This seems justified by the fact that the expression obtained in the end for the large deviation function is analytic in $u$, and by comparison with numerical evaluations of the probability density at small times, see figure \ref{Fig LDF}.

\begin{subsection}{Case \texorpdfstring{$u>0$}{u>0}}
\label{appendix saddle point u>0}
We consider here that $z$ and $d$, $\Re\,d>0$ are such that it is possible to choose a determination of the logarithm so that its branch cut is never crossed in (\ref{h[int w]}). Taking the derivative with respect to $d$ of (\ref{h[int w]}) and using $\log(1+q^{-1})=\log(1+q)-\log q$ for $q\not\in\mathbb{R}^{-}$, one finds for generic values of $d$
\begin{equation}
\fl\hspace{15mm} \partial_{d}h(u,d,z)=\frac{2(4\pi d)^{3/2}}{3}-2u\sqrt{4\pi d}+\pi+\rmi\log(-z\,\rme^{\rmi\pi+\frac{2\rmi}{3}(4\pi d)^{3/2}-2\rmi u\sqrt{4\pi d}})\;.
\end{equation}
Writing $z=r\,\rme^{\rmi\theta}$, $r>0$, $\theta\in\mathbb{R}$, it leads to
\begin{equation}
\fl\hspace{15mm} \partial_{d}h(u,d,z)=\rmi\log(-z)+2\pi\Big\lfloor\frac{2\pi+\theta+\Im(\frac{2\rmi}{3}(4\pi d)^{3/2}-2\rmi u\sqrt{4\pi d})}{2\pi}\Big\rfloor\;,
\end{equation}
with $\lfloor x\rfloor$ the largest integer lower than $x$. We define a domain $\mathcal{D}_{u}\subset\mathbb{C}$, equal to (the unbounded connected component of)
\begin{equation}
\label{Du Red>0}
\mathcal{D}_{u}=\{d,\Re\,d>0,\Big\lfloor\frac{2\pi+\Im(\frac{2\rmi}{3}(4\pi d)^{3/2}-2\rmi u\sqrt{4\pi d})}{2\pi}\Big\rfloor=0\}\;.
\end{equation}
The domain $\mathcal{D}_{u}$ always contains a suitable path between $\rme^{\rmi\pi/3}\infty$ and $\rme^{-\rmi\pi/3}\infty$, see figure \ref{Fig domain D}.

For $d\in\mathcal{D}_{u}$ and $z$ in a neighbouring of $-1$, one has $\partial_{d}h(u,d,z)=\rmi\log(-z)$, hence the function $\Xi(u)=h(u,d,-1)$ is independent of $d$. Its expression obtained from (\ref{h[int w]}) is equal to the sum of two integrals with the same integrand, and contours of integration $\Gamma_{+}$ and $\Gamma_{-}$. Since $\Re\,d>0$, the finite ends $\alpha_{+}$ and $\alpha_{-}$ of the contours verify $\alpha_{+}+\alpha_{-}=0$.

\begin{figure}
  \begin{center}
    \begin{tabular}{c}
      \includegraphics[width=100mm]{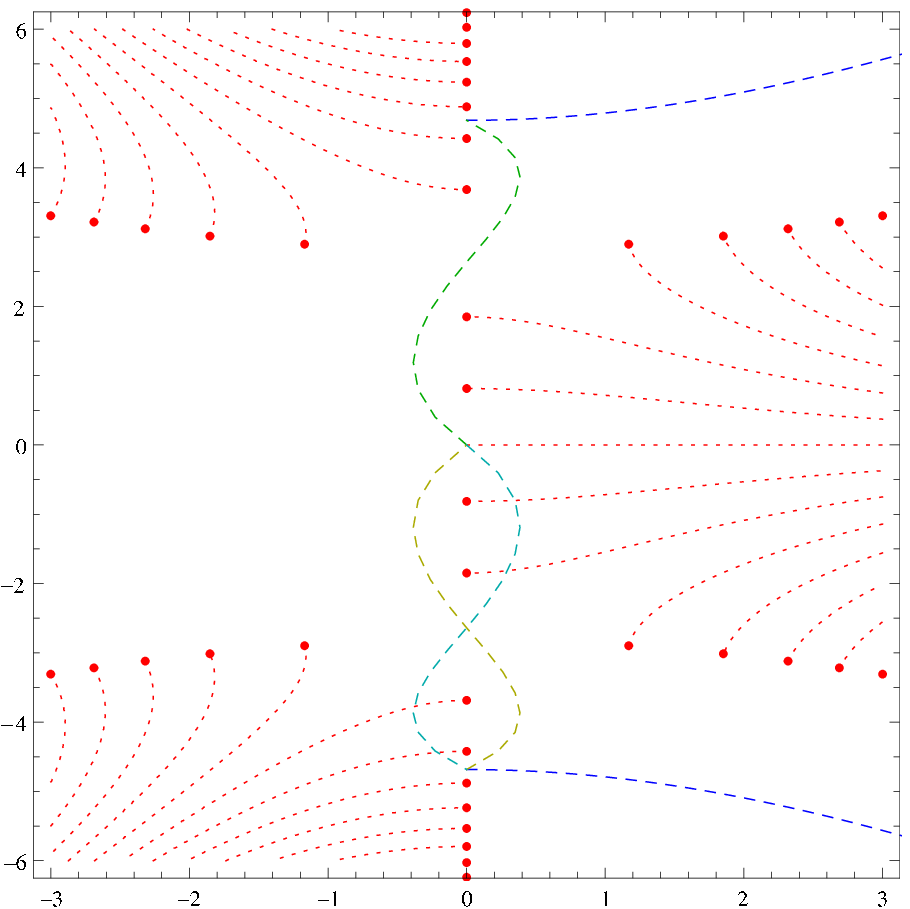}
      \begin{picture}(0,0)
        \put(-30,91){\small\color{blue}$\Gamma_{+}$}
        \put(-30,9){\small\color{blue}$\Gamma_{-}$}
        \put(-60,67){\small\color[rgb]{0,0.6,0}$\Lambda_{+}$}
        \put(-62,40){\small\color[rgb]{0.6,0.6,0}$\Lambda_{-}$}
        \put(-45,34){\small\color[rgb]{0,0.6,0.6}$-\Lambda_{+}$}
      \end{picture}
    \end{tabular}
  \end{center}
  \caption{Contours $\Gamma_{\pm}$, $\Lambda_{+}$, $\Lambda_{-}$ and $-\Lambda_{+}$ (blue, green, yellow, cyan, dashed lines) for $u=5$ and $d=1.1\in\mathcal{D}_{u}$, plotted along with the branch cuts (red, dotted lines) and poles (red dots; after integration by parts) in the variable $w$ of the integrand in (\ref{identity close contour Xi Red>0}). All contours are oriented upward except $-\Lambda_{+}$ that is oriented downward.}
  \label{Fig contours Gamma Lambda Red>0}
\end{figure}
The expression of $\Xi(u)$ from (\ref{h[int w]}) can be simplified by closing the contour between $\alpha_{-}$ and $\alpha_{+}$ on the path $\Lambda_{+}\cup\Lambda_{-}$ represented in figure \ref{Fig contours Gamma Lambda Red>0} for generic $u>0$ and $d\in\mathcal{D}_{u}\cap\mathbb{R}^{+}$ (which implies that $\alpha_{\pm}$ is purely imaginary, and the contours $\Gamma_{\pm}$ stay in the half-plane with positive real part; if $d\not\in\mathbb{R}^{+}$, either $\Im\,\alpha_{+}<0$ or $\Im\,\alpha_{-}<0$ and the branch cuts with $\Re\,w<0$ have then to be chosen in such a way that they do not intersect the contours $\Gamma_{\pm}$). After an integration by parts to replace branch cuts with poles, using again $\log(1+q^{-1})=\log(1+q)-\log q$ (for generic $q$) and (\ref{Du Red>0}), one finds
\begin{eqnarray}
\label{int Lambda+-}
&& \int_{\Lambda_{-}\cup\Lambda_{+}}\frac{w\,\rmd w}{2^{5/3}\rmi\pi}\,\log\Big(1-\rme^{\frac{w^{3}}{3}+2^{2/3}uw}\Big)
=\frac{(4\pi d)^{5/2}}{6\pi}-u\,\frac{(4\pi d)^{3/2}}{2\pi}+\pi d\\
&&\hspace{68mm} -\int_{\Lambda_{-}\cup\Lambda_{+}}\frac{w^{2}\,\rmd w}{2^{8/3}\rmi\pi}\,\frac{w^{2}+2^{2/3}u}{1-\rme^{-\frac{w^{3}}{3}-2^{2/3}uw}}\;.\nonumber
\end{eqnarray}
All branch points have become poles except $w=0$ for which the residue vanishes. Replacing $w$ by $-w$ in the integral on the contour $\Lambda_{+}$ and using $(1-\rme^{\frac{w^{3}}{3}+2^{2/3}uw})^{-1}=1-(1-\rme^{-\frac{w^{3}}{3}-2^{2/3}uw})^{-1}$, the integral in the right hand side of (\ref{int Lambda+-}) becomes an integral over the closed path $\Lambda_{-}\cup(-\Lambda_{+})$, see figure \ref{Fig contours Gamma Lambda Red>0}. We distinguish two kinds of poles inside this contour: "inner poles" $w_{k}^{\text{in}}$, $k=1,\ldots,n$, located closest from $0$, that the contour encircles clockwise, and "outer poles" $w_{k}^{\text{out}}$, $k=1,\ldots,n$, located farthest from $0$, that the contour encircles counter-clockwise. The integer $n$ is equal to the number of strictly positive $k$ such that $\frac{w_{k}^{3}}{3}+2^{2/3}uw_{k}=-2\rmi\pi k$ has $3$ real roots $\rmi w_{k}$. This is equivalent to $\Delta=16u^{3}/3-3(2\pi k)^{2}>0$, where $\Delta$ is the discriminant of the third degree equation. One has thus $n=\lfloor\frac{2u^{3/2}}{3\pi}\rfloor$. The residues of inner and outer poles are equal to
\begin{eqnarray}
\underset{w=w_{k}^{\text{in}}}{\Res}\,\frac{w^{2}}{2^{5/3}}\,\frac{w^{2}+2^{2/3}u}{1-\rme^{-\frac{w^{3}}{3}-2^{2/3}uw}}=-2u\sin^{2}\!\Big(\frac{\arcsin\frac{3\pi k}{2u^{3/2}}}{3}\Big)\\
\underset{w=w_{k}^{\text{out}}}{\Res}\,\frac{w^{2}}{2^{5/3}}\,\frac{w^{2}+2^{2/3}u}{1-\rme^{-\frac{w^{3}}{3}-2^{2/3}uw}}=-2u\sin^{2}\!\Big(\frac{-\pi+\arcsin\frac{3\pi k}{2u^{3/2}}}{3}\Big)\;.
\end{eqnarray}
One finally finds that the integral on the path from $\alpha_{-}$ to $\alpha_{+}$ is equal to
\begin{equation}
\label{identity close contour Xi Red>0}
\fl\hspace{15mm} \int_{\alpha_{-}}^{\alpha_{+}}\frac{w\,\rmd w}{2^{5/3}\rmi\pi}\,\log\Big(1-\rme^{\frac{w^{3}}{3}+2^{2/3}uw}\Big)
=\frac{(4\pi d)^{5/2}}{15\pi}-u\,\frac{(4\pi d)^{3/2}}{3\pi}+\pi d-\text{a.c.}(u)\;,
\end{equation}
with
\begin{equation}
\label{a.c.+(u)}
\fl\hspace{20mm} \text{a.c.}(u)=2u\sum_{k=1}^{\lfloor\frac{2u^{3/2}}{3\pi}\rfloor}\bigg(\sin^{2}\Big(\frac{\arcsin\frac{3\pi k}{2u^{3/2}}}{3}\Big)-\sin^{2}\Big(\frac{-\pi+\arcsin\frac{3\pi k}{2u^{3/2}}}{3}\Big)\bigg)\;.
\end{equation}
This leads to the integral expression in (\ref{Xi[int,Ai]}) for the function $\Xi(u)$. We observe that the additional term $\text{a.c.}(u)$ is precisely the one required to make $\Xi$ analytic near the positive real axis.

The function $\Xi(u)$ can also be expressed in terms of the Airy function by expanding the logarithm and using
\begin{equation}
\Ai(q)=\int_{-\rmi\infty+\epsilon}^{\rmi\infty+\epsilon}\frac{\rmd w}{2\rmi\pi}\,\rme^{\frac{w^{3}}{3}-qw}\;,
\end{equation}
with $\epsilon>0$. This leads to the expression with the infinite sum of $\Xi$ in (\ref{Xi[int,Ai]}).
\end{subsection}

\begin{subsection}{Case \texorpdfstring{$u<0$}{u<0}}
\label{appendix saddle point u<0}
\begin{figure}
  \begin{center}
    \begin{tabular}{c}
      \includegraphics[width=100mm]{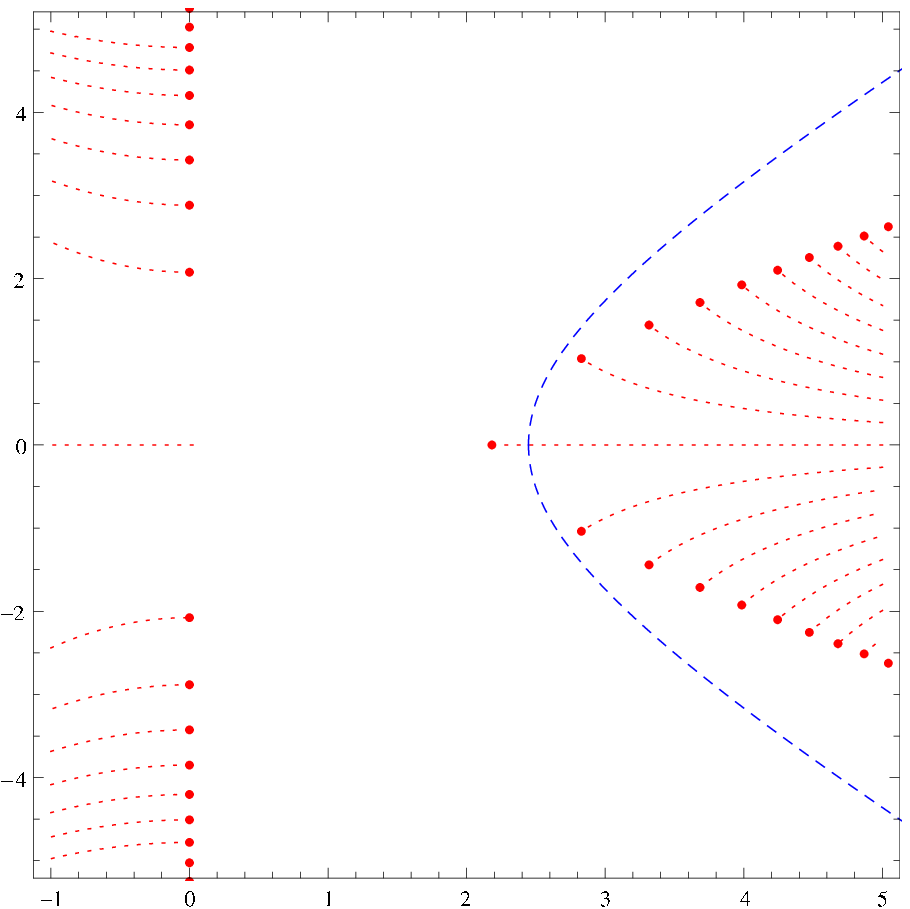}
      \begin{picture}(0,0)
        \put(-20,85){\small\color{blue}$\Gamma_{+}$}
        \put(-20,16){\small\color{blue}$\Gamma_{-}$}
      \end{picture}
    \end{tabular}
  \end{center}
  \caption{Contour $\Gamma=\Gamma_{+}\cup\Gamma_{-}$ (blue, dashed line) for $u=-1$ and $d=-0.3\in\mathcal{D}_{u}$, plotted along with the branch cuts (red, dotted lines) and poles (red dots; after integration by parts) in the variable $w$ of the integrand in (\ref{h(u,d,-1) Red<0 single integral}). The contour is oriented upward.}
  \label{Fig contour Gamma Red<0}
\end{figure}
We consider here $d$ in the half-plane $\Re\,d<0$. It implies that the finite extremities $\alpha_{\pm}$ of the contours $\Gamma_{\pm}$ are equal. The expression (\ref{h[int w]}) of $h(u,d,-1)$ can then be written in terms of a single integral with a connected path of integration $\Gamma=\Gamma_{+}\cup\Gamma_{-}$ between $\sqrt{-\rmi}\,\infty$ and $\sqrt{\rmi}\,\infty$:
\begin{equation}
\label{h(u,d,-1) Red<0 single integral}
h(u,d,-1)=2\pi d+\int_{\Gamma}\frac{w\,\rmd w}{2^{5/3}\rmi\pi}\,\log\Big(1-\rme^{\frac{w^{3}}{3}+2^{2/3}uw}\Big)\;.
\end{equation}

We introduce the domain
\begin{equation}
\label{Du Red<0}
\mathcal{D}_{u}=\Big\{d,\frac{3u}{8\pi}-\frac{3\sqrt{2}}{16\sqrt{3u-4\pi\Re\,d}}<\Re\,d<\frac{3u}{4\pi}\Big\}\;,
\end{equation}
which is a vertical strip in the complex plane, see figure \ref{Fig domain D}. Using $\Re(w^{3})=(\Re\,w)^{3}-3(\Re\,w)(\Im\,w)^{2}$, $\Im(w^{3})=-(\Im\,w)^{3}+3(\Re\,w)^{2}(\Im\,w)$ and $(\Re\,w)^{2}=(\Im\,w)^{2}-2^{8/3}\pi\,\Re\,d$ for $w\in\Gamma$, one can show that for $d\in\mathcal{D}_{u}$, the contour $\Gamma$ crosses the real branch cut (an only that branch cut) of the integrand in the expression (\ref{h(u,d,-1) Red<0 single integral}) of $h(u,d,-1)$, see figure \ref{Fig contour Gamma Red<0}.

Integrating (\ref{h(u,d,-1) Red<0 single integral}) by parts, we observe that the contributions of both sides of the branch cut cancel the term $2\pi d$. We obtain
\begin{equation}
h(u,d,-1)=-\int_{\Gamma}\frac{w^{2}\,\rmd w}{2^{8/3}\rmi\pi}\,\frac{w^{2}+2^{2/3}u}{1-\rme^{-\frac{w^{3}}{3}-2^{2/3}uw}}\;.
\end{equation}
The point $z=-1$ is again a saddle point with respect to $d$, $\partial_{d}h(u,d,-1)=0$. Moving the contour to the left, one picks the term $3u/2$ from the residue of the pole $w_{0}=2^{1/3}\sqrt{3}\sqrt{-u}$. Integrating by parts again, we recover (\ref{Xi[int,Ai]}) with a contour of integration that does not cross any branch cut. Again, we observe that the extra term $3u/2$ corresponds precisely to the analytic continuation to $u<0$.
\end{subsection}
\end{section}


\end{document}